\documentclass[twocolumn, journal]{IEEEtran}
\IEEEoverridecommandlockouts

\usepackage{color}
\usepackage{bm}
\usepackage{graphicx}
\usepackage{amsmath}
\usepackage{amssymb}
\usepackage{algorithm}
\usepackage{algorithmic}
\usepackage{multirow}
\usepackage{booktabs}
\usepackage{array}
\usepackage{amsthm}
\usepackage{lipsum}
\usepackage{enumerate}
\usepackage{stfloats}
\usepackage{subfigure}
\usepackage{cases}
\usepackage{diagbox}
\usepackage{threeparttable}
\usepackage{cite}

\newcommand{\non}{\nonumber}

\title{Synergizing Beyond Diagonal \\Reconfigurable Intelligent Surface and Rate-Splitting Multiple Access}

\author{Hongyu Li,~\IEEEmembership{Graduate Student Member,~IEEE}, Shanpu Shen,~\IEEEmembership{Senior Member,~IEEE},\\ and Bruno Clerckx,~\IEEEmembership{Fellow,~IEEE}
\thanks{Manuscript received; \textit{(Corresponding author: Shanpu Shen).}}
\thanks{Hongyu Li is with the Department of Electrical and Electronic Engineering, Imperial College London, London SW7 2AZ, U.K. (e-mail: c.li21@imperial.ac.uk).}
\thanks{Shanpu Shen is with the Department of Electrical Engineering and Electronics, University of Liverpool, Liverpool L69 3GJ, U.K. (email: Shanpu.Shen@liverpool.ac.uk).}
\thanks{Bruno Clerckx is with the Department of Electrical and Electronic Engineering, Imperial College London, London SW7 2AZ, U.K. and with Silicon Austria Labs (SAL), Graz A-8010, Austria (e-mail: b.clerckx@imperial.ac.uk; bruno.clerckx@silicon-austria.com).}}

\begin{document}

\maketitle
\thispagestyle{empty}
\begin{abstract}
This work focuses on the synergy of rate-splitting multiple access (RSMA) and beyond diagonal reconfigurable intelligent surface (BD-RIS) to enlarge the coverage, improve the performance, and save on antennas. 
Specifically, we employ a multi-sector BD-RIS modeled as a prism, which can achieve highly directional full-space coverage, in a multiuser multiple input single output communication system.   
With the multi-sector BD-RIS aided RSMA model, we jointly design the transmit precoder and BD-RIS matrix under the imperfect channel state information (CSI) conditions. 
The robust design is performed by solving a stochastic average sum-rate maximization problem. 
With sample average approximation and weighted minimum mean square error-rate relationship, the stochastic problem is transformed into a deterministic one with multiple blocks, each of which is iteratively designed. 
Simulation results show that multi-sector BD-RIS aided RSMA outperforms space division multiple access schemes. More importantly, synergizing multi-sector BD-RIS with RSMA is an efficient strategy to reduce the number of active antennas at the transmitter and the number of passive antennas in BD-RIS. 
\end{abstract}


\section{Introduction}

The next generation of wireless communications is expected to improve ongoing techniques in 5G and explore novel techniques to support high data rate transmission and massive connectivity \cite{tataria20216g}.
Specifically in the physical (PHY) layer, rate-splitting multiple access (RSMA) \cite{mao2022rate,clerckx2023primer} and reconfigurable intelligent surface (RIS) \cite{di2020smart} are two promising techniques, which can improve the wireless communication performance, such as spectral efficiency, energy efficiency, and coverage.

RSMA is a PHY-layer non-orthogonal transmission strategy, which enables more flexible inter-user interference management compared to other existing multiple access techniques, such as spatial division multiple access (SDMA) and orthogonal multiple access (OMA) \cite{mao2022rate}. The key principle of (1-layer) RSMA is summarized as follows. 
1) At the transmitter, user messages are split into common and private parts, where the common parts from different users are combined and encoded into one common stream, while the private parts are individually encoded into private streams. 
2) At the receiver, each user first decodes the common stream and performs successive interference cancellation (SIC) to remove the common stream, and then decodes its own private stream. 
Such a technique has been proved to achieve higher degree of freedom (DoF) \cite{joudeh2016sum}, spectral and energy efficiency \cite{mao2019rate} than SDMA, and be robust to channel state information (CSI) uncertainty \cite{joudeh2016sum} and mobility \cite{dizdar2021rate}. 
The higher efficiency, universality, flexibility, robustness, and reliability of RSMA over existing multiple access strategies have been demonstrated in more than forty different applications and scenarios relevant to 6G \cite{mao2022rate,clerckx2023primer}.

Meanwhile, RIS is an emerging PHY-layer technique, which enables flexible manipulations of wireless communication environments in an energy-efficient manner \cite{di2020smart}. The key advantages of RIS come from the following two perspectives. 
1) RIS is a planar surface consisting of a large number of nearly-passive elements with low power consumption. 
2) RIS has tuneable elements which can reconfigure incident signals. 
Such a technique has been widely employed into various schemes, such as wireless power transfer (WPT) \cite{feng2022waveform}, simultaneous wireless information and power transfer (SWIPT) \cite{zhao2021irs}, integrated sensing and communication (ISAC) \cite{liu2022joint}, to show the benefits of RIS in improving sum output current \cite{feng2022waveform,zhao2021irs} and radar output signal-to-interference-plus-noise (SINR) \cite{liu2022joint}.

% \begin{figure}
%     \centering
%     \includegraphics[width=0.48\textwidth]{figures/RIS_Tree.eps}
%     \caption{RIS classification tree.}
%     \label{fig:RIS_tree}
% \end{figure}

Conventional RIS utilized in aforementioned research \cite{feng2022waveform,zhao2021irs,liu2022joint} is modeled as a \textit{diagonal} phase shift matrix, assuming an architecture where each RIS element is connected to ground by its own load without interacting with other elements. 
However, new RIS architectures have recently appeared, namely beyond diagonal RIS (BD-RIS), whose mathematical model is not limited to diagonal matrices. 
Until now, there are three categories of BD-RIS.
The \textit{first} category is BD-RIS with group/fully-connected architectures modeled as block diagonal matrices \cite{shen2021,nerini2021reconfigurable,nerini2022optimal,zhang2022intelligent,li2022,li2022beyond}. In this category, group/fully-connected BD-RIS with reflective mode is first proposed and modeled in \cite{shen2021}, followed by the study of discrete-value group/fully-connected architectures \cite{nerini2021reconfigurable} and optimal group/fully-connected design \cite{nerini2022optimal} of BD-RIS. 
To enlarge the coverage of RIS, simultaneous transmitting and reflecting (STAR) RIS or intelligent omni-surface (IOS) \cite{zhang2022intelligent} is recently introduced and implemented. Then STAR-RIS/IOS is proved to be a special case of group-connected architecture with group size 2 and further generalized in \cite{li2022}. 
Multi-sector BD-RIS is proposed in \cite{li2022beyond} to further enhance the achievable performance of RIS while guaranteeing the full-space coverage.
The \textit{second} category is BD-RIS with dynamically group-connected architectures modeled as permuted block diagonal matrix \cite{li2022dynamic}. 
Different from the first category, where the architectures are fixed regardless of CSI, a dynamically group-connected architecture adapting to CSI is proposed in \cite{li2022dynamic} based on a dynamic grouping strategy to further enhance the performance. 
The \textit{third} category is BD-RIS with non-diagonal phase shift matrix \cite{li2022reconfigurable}. 
In this category, a novel architecture, where signals impinging on one element can be reflected from another element, is modeled, analyzed, and proved to achieve better performance than conventional RIS with diagonal phase shift matrices.

Due to the potential advantages of RSMA and RIS, the synergy among them has recently gained much attention \cite{de2022rate,li2022rate}. 
The synergy among the conventional RIS and 1-layer RSMA in multiuser systems is most commonly considered, whose benefits are shown by focusing on various metrics, such as sum-rate maximization \cite{wu2023deep}, energy efficiency maximization \cite{yang2020energy,weinberger2022synergistic}, outage probability analysis \cite{bansal2021rate,shambharkar2022rate,bansal2022rate}, transmit power minimization \cite{weinberger2021synergistic}, and max-min fairness \cite{zhao2022reconfigurable}.
In addition, the interplay between conventional RIS and RSMA beyond 1-layer architectures, such as 2-layer hierarchical RSMA \cite{bansal2021analysis,jolly2021analysis}, is also studied. 
While the abovementioned RIS aided RSMA work \cite{yang2020energy,weinberger2022synergistic,bansal2021rate,shambharkar2022rate,bansal2022rate,weinberger2021synergistic,zhao2022reconfigurable,bansal2021analysis,jolly2021analysis} is restricted to perfect CSI conditions, a more practical imperfect CSI scenario is analyzed in \cite{weinberger2022sacrificing}. 
However, the above results are restricted to the integration of RSMA and conventional RIS, which limits both the coverage and the achievable performance. 
To realize full-space coverage in wireless networks, there are a few results on STAR-RIS aided RSMA considering spatially correlated channels \cite{dhok2022rate}, hardware impairments \cite{soleymani2022rate}, PHY layer security \cite{hashempour2022secure}, and the coupling of transmission and reflection phases \cite{katwe2023improved}. 
Meanwhile, to enhance the spectrum efficiency of wireless communications, there is only one work \cite{fang2022fully} considering the employment of BD-RIS with fully-connected architecture and reflective mode in RSMA transmission network, where the sum-rate enhancement compared to SDMA schemes is highlighted. 
Nevertheless, none of the existing work investigates the integration of BD-RIS and RSMA to simultaneously 1) enlarge the coverage, 2) boost the rate performance \cite{fang2022fully}, and 3) reduce the required number of active/passive antennas to improve energy efficiency, which motivates this work.

The contributions of this work are summarized as follows. 

\textit{First}, we consider the integration of RSMA and multi-sector BD-RIS in a multiuser multiple input single output (MU-MISO) system to simultaneously achieve the abovementioned three goals.
The first two goals are achieved by adopting the multi-sector BD-RIS, which is modeled as a prism consisting of multiple sectors with each covering part of the horizontal space. More importantly, the multi-sector BD-RIS enables highly directional beams and thus can provide high performance gains, thanks to the use of high-gain antennas with narrow beamwidth.  
To achieve the third goal, we adopt 1-layer RSMA at the transmitter, which provides additional DoF over conventional multiuser linear precoding, enables better inter-user interference management, and has the potential to save on the number of both active and passive antennas when synergizing with BD-RIS. 

\textit{Second}, we propose a beamforming design algorithm to maximize the average sum-rate for multi-sector BD-RIS aided RSMA under imperfect CSI conditions. To solve the stochastic problem, we first approximate it as a deterministic optimization by sample average approximation (SAA) methods \cite{shapiro2021lectures}, and then transform the deterministic problem into a more tractable three-block optimization based on the weighted minimum mean square error (WMMSE)-rate relationship \cite{christensen2008weighted}. Specifically for the design of the BD-RIS matrix block, due to the newly introduced constraint of the multi-sector BD-RIS, the approaches for conventional RIS cases are not feasible.  Therefore, we propose a novel algorithm by first transforming the non-smooth objective into a smooth one, and then solving the smooth problem by manifold methods.
Initialization, complexity analysis, and convergence analysis of the proposed design are also provided.

\textit{Third}, we present simulation results to evaluate the ergodic sum-rate performance of the multi-sector BD-RIS aided RSMA compared to the SDMA scheme from the perspective of 1) the radiation pattern of BD-RIS antennas, 2) CSI conditions, and 3) the quality of service (QoS) threshold. Simulation results show that multi-sector BD-RIS aided RSMA always outperforms multi-sector BD-RIS aided SDMA in either of the abovementioned perspectives.

\textit{Fourth}, with the same ergodic sum-rate requirement and QoS thresholds, we show the required numbers of active and passive antennas can be reduced compared to the SDMA scheme by synergizing BD-RIS with RSMA. Such reduction of antenna numbers leads to a potential energy efficiency gain.

\textit{Organization:} Section II illustrates the multi-sector BD-RIS aided RSMA model. 
Section III provides the robust beamforming design for the multi-sector BD-RIS aided RSMA.
Section IV evaluates the performance of the proposed design and Section V concludes this work.

\textit{Notations}:
Boldface lower- and upper-case letters indicate column vectors and matrices, respectively.
$\mathbb{C}$, $\mathbb{R}$, and $\mathbb{R}^+$ denote the set of complex numbers, real numbers, and non-negative real numbers, respectively.
$\mathbb{E}\{\cdot\}$ represents statistical expectation.
$(\cdot)^\ast$, $(\cdot)^T$, $(\cdot)^H$, and $(\cdot)^{-1}$ denote the conjugate, transpose, conjugate-transpose operations, and inversion, respectively.
$\Re\{\cdot\}$ denotes the real part of a complex number.
$\mathbf{I}_L$ denotes an $L \times L$ identity matrix.
$\mathbf{0}$ denotes an all-zero matrix.
$\mathcal{CN}(\mathbf{0},\mathbf{\Sigma})$ denotes the circularly symmetric complex Gaussian (CSCG) distribution with mean $\mathbf{0}$ and covariance matrix $\mathbf{\Sigma}$.
$\|\mathbf{A}\|_F$ denotes the Frobenius norm of matrix $\mathbf{A}$.
$\|\mathbf{a}\|_2$ denotes the $\ell_2$-norm of vector $\mathbf{a}$.
$|a|$ denotes the norm of variable $a$.
$\mathsf{vec}(\cdot)$ denotes the vectorization of a matrix.
$\mathsf{diag}(\cdot)$ denotes a diagonal matrix.
$\mathsf{Tr}\{\cdot\}$ denotes the summation of diagonal elements of a matrix. 
$\triangledown(\cdot)$ denotes the first-order Euclidean gradient operation. 
$\frac{\partial(\cdot)}{\partial(\cdot)}$ denotes the first-order partial gradient operation.
$[\mathbf{A}]_{i,:}$, $[\mathbf{A}]_{:,j}$, $[\mathbf{A}]_{i,j}$, and $[\mathbf{a}]_i$ denote the $i$-th row, the $j$-th column, the $(i,j)$-th element of matrix $\mathbf{A}$, and the $i$-th element of vector $\mathbf{a}$, respectively.

\section{Multi-Sector BD-RIS Aided RSMA}
\label{sec:BD-RIS_RSMA}
In this section, we briefly review the concept of multi-sector BD-RIS, describe the multi-sector BD-RIS aided MU-MISO system, explain the architecture of 1-layer RSMA, and illustrate the modeling of CSI uncertainty.

\begin{figure}
    \centering
    \includegraphics[width=0.48\textwidth]{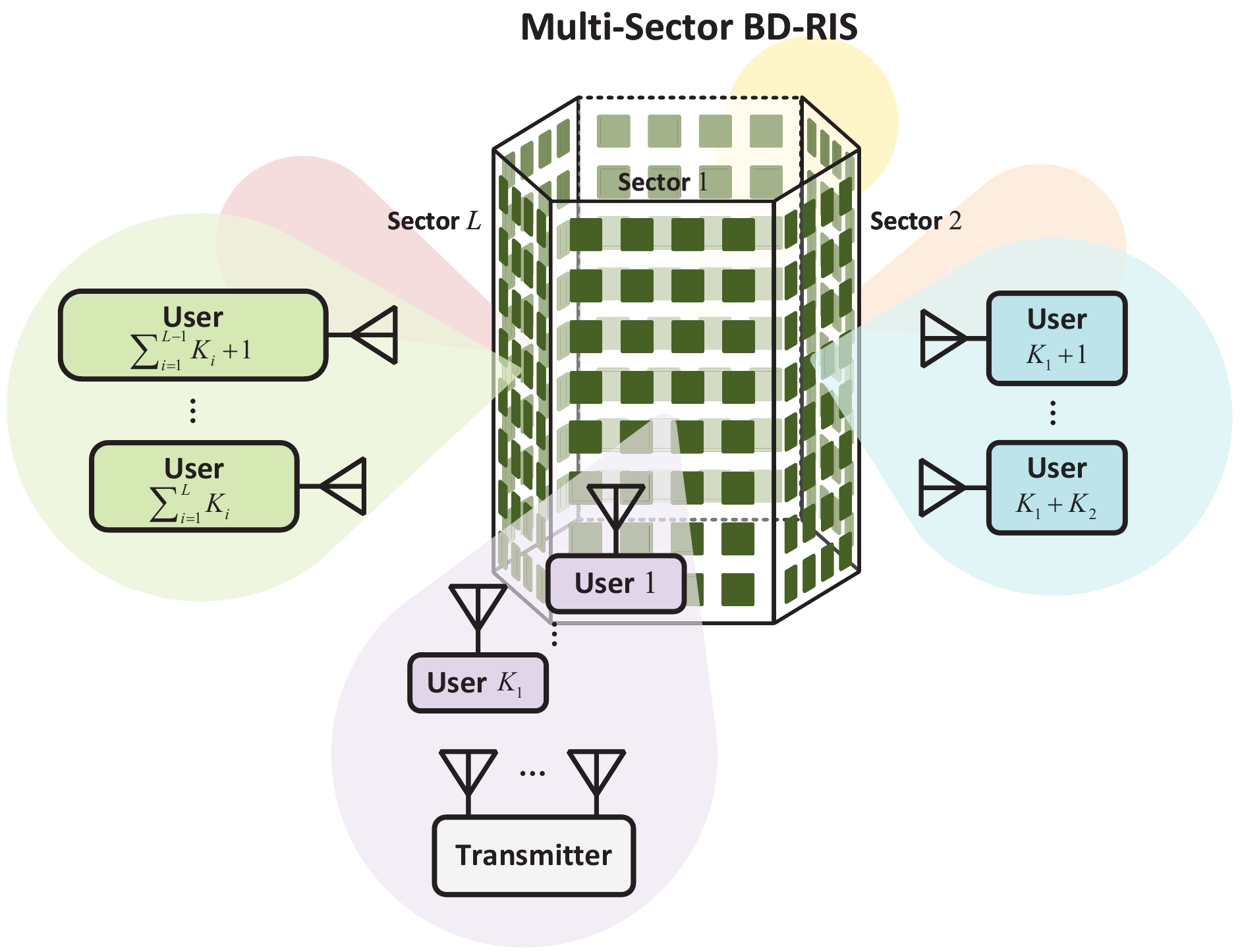}
    \caption{A paradigm of a multi-sector BD-RIS aided MU-MISO system.}\vspace{-0.2 cm}
    \label{fig:syst_mod}
\end{figure}

\subsection{Multi-Sector BD-RIS}
\label{sec:BD-RIS_review}
The multi-sector BD-RIS is modeled as a polygon consisting of $L$ sectors, each of which has $M$ antennas, as shown in Fig. \ref{fig:syst_mod}. 
The multi-sector BD-RIS is modeled as multiple antennas connected to a group-connected reconfigurable impedance network \cite{li2022beyond}. 
In this subsection, we briefly explain the key of multi-sector BD-RIS from the following two perspectives.

1) The multi-sector BD-RIS consists of $M$ cells with an index set $\mathcal{M} = \{1,\ldots,M\}$. Particularly, cell $m$, $\forall m\in\mathcal{M}$ contains antennas $m$, $M+m$, $\dots$, $(L-1)M+m$, which are connected to each other by reconfigurable impedance components to support the multi-sector mode.

2) Each multi-sector BD-RIS antenna has a uni-directional radiation pattern with beamwidth $2\pi/L$, which only covers $1/L$ azimuth space\footnote{In practice, the uni-directional radiation pattern can be realized by patch antennas.}. 
In each cell, $L$ antennas are located at the edge of the polygon such that $L$ antennas cover the whole azimuth space. 
The larger the number of sectors $L$, the narrower the beamwidth of each BD-RIS antenna, and thus the higher the gain of each BD-RIS antenna.

To summarize, benefiting from group-connected impedance network and antenna array arrangements, the multi-sector BD-RIS can realize enlarged wireless coverage, more flexible wave manipulation, and enhanced channel gain compared to conventional RIS. 

Mathematically, the multi-sector BD-RIS is characterized by $L$ matrices, $\mathbf{\Phi}_l\in\mathbb{C}^{M\times M}$, $\forall l\in\mathcal{L}=\{1,\ldots,L\}$, corresponding to $L$ sectors, which are sub-matrices of the scattering matrix $\mathbf{\Phi}\in\mathbb{C}^{LM\times LM}$ for the $LM$-port reconfigurable impedance network, i.e., $\mathbf{\Phi}_l = [\mathbf{\Phi}]_{(l-1)M+1:lM,1:M}$.
They satisfy a combined unitary constraint given by \cite{li2022beyond}
\begin{equation}
    \sum\nolimits_{l}\mathbf{\Phi}_l^H\mathbf{\Phi}_l = \mathbf{I}_M.
    \label{eq:bd_ris}
\end{equation}
More details about the modeling, design, and performance evaluation of the multi-sector BD-RIS can be found in \cite{li2022beyond}.

\subsection{Multi-Sector BD-RIS Aided MU-MISO}
Consider a MU-MISO system operating in the downlink as illustrated in Fig. \ref{fig:syst_mod}, where the transmitter equipped with $N$ antennas communicates to a set of single-antenna users $\mathcal{K} = \{1,\ldots,K\}$ with the aid of an $M$-cell multi-sector BD-RIS\footnote{In this work, we ignore the direct links between the transmitter and users for simplicity, while the proposed design in the following section is still feasible when direct links exist.}.
In this work, we assume the transmitter is within the coverage of sector 1 of the multi-sector BD-RIS and $K_l$ users in $\mathcal{K}_l = \{\sum_{i=1}^{l-1}K_i+1,\sum_{i=1}^{l-1}K_i+2,\ldots,\sum_{i=1}^{l}K_i\}$ are located within the coverage of sector $l$, $\forall l\in\mathcal{L}$ of the multi-sector BD-RIS\footnote{In this work, we consider the scenario that the relative location between each user and each sector of BD-RIS is fixed for simplicity such that there is no user selection issue. We also assume the sidelobe of each BD-RIS antenna is negligible compared to the mainlobe such that both transmitter and each user are covered by only one sector of the BD-RIS.}, $\cup_{l}\mathcal{K}_l = \mathcal{K}$, as illustrated in Fig. \ref{fig:syst_mod}.  
For the multi-sector BD-RIS aided downlink MU-MISO, the signal received by user $k$, $\forall k\in\mathcal{K}_l$ covered by sector $l$, $\forall l\in\mathcal{L}$ is 
\begin{equation}
    y_k = \mathbf{h}_k^H\mathbf{\Phi}_l\mathbf{G}\mathbf{x} + n_k, 
    \label{eq:received_signal}
\end{equation}
where $\mathbf{h}_k\in\mathbb{C}^M$ and $\mathbf{G}\in\mathbb{C}^{M\times N}$, respectively, are the channel between the multi-sector BD-RIS and user $k$, $\forall k\in\mathcal{K}$, and the channel between the transmitter and the BD-RIS, $\mathbf{x}\in\mathbb{C}^N$ is the transmit signal vector, and $n_k\sim\mathcal{CN}(0,\sigma_k^2)$, $\forall k\in\mathcal{K}$ is the additive white Gaussian noise (AWGN). 
The transmit signal vector $\mathbf{x}$ is subject to the power constraint $\mathbb{E}\{\mathbf{x}^H\mathbf{x}\}\le P$, where $P$ is the maximum transmit power.

\begin{figure}
    \centering
    \includegraphics[width=0.48\textwidth]{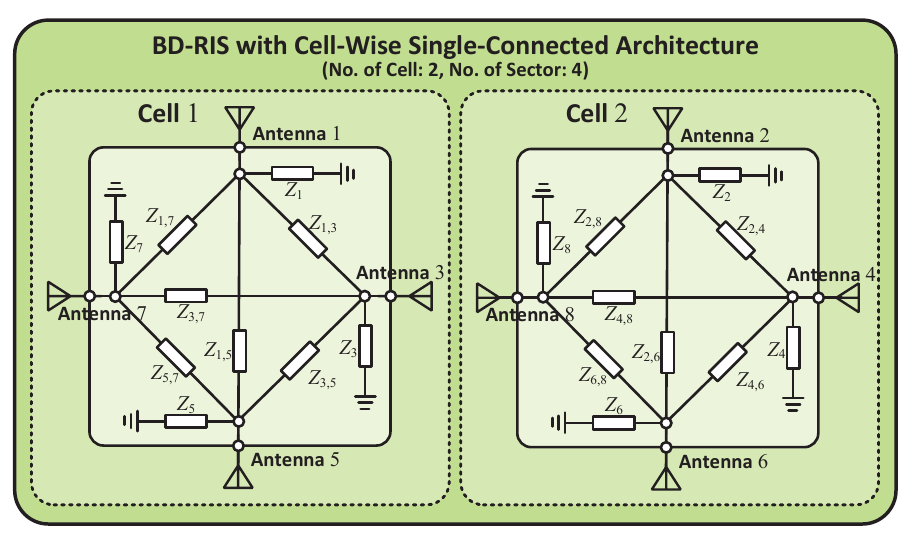}
    \caption{Diagram of cell-wise single-connected BD-RIS with 4 sectors and 2 cells, each of which has 4 antennas.}\vspace{-0.2 cm}
    \label{fig:RIS}
\end{figure}

As discussed in \cite{shen2021,li2022,li2022beyond}, reconfigurable impedance networks with different circuit topologies have scattering matrices with different mathematical characteristics. 
In this work, we focus on a cell-wise single-connected (CW-SC) architecture of the multi-sector BD-RIS, where the inner-cell antennas are connected to each other by reconfigurable impedance components while the inter-cell antennas are not connected to each other\footnote{The reconfigurable impedance components to construct the circuit topology can be realized by using tunable inductance and capacitance, such as varactors with continuous values or PIN diodes with discrete states.}. 
To facilitate understanding, we provide an example of the BD-RIS with 4 sectors, 2 cells, and the CW-SC architecture in Fig. \ref{fig:RIS}.
The CW-SC architecture of the multi-sector BD-RIS results in diagonal matrices of $\mathbf{\Phi}_l$, $\forall l\in\mathcal{L}$, each of which can be described as $\mathbf{\Phi}_l = \mathsf{diag}(\phi_{l,1},\ldots,\phi_{l,M})$ with $\phi_{l,m}\in\mathbb{C}$, $\forall l\in\mathcal{L}$, $\forall m\in\mathcal{M}$. 
Therefore, the constraint (\ref{eq:bd_ris}) can be re-written as
\begin{equation}
    \label{eq:bd_ris_sc}
    \sum\nolimits_{l} |\phi_{l,m}|^2 = 1, \forall m,
\end{equation}
which indicates that the transmit wave impinging on one sector of BD-RIS is flexibly split and reflected by other sectors, where the proportion of energy split to different sectors is determined by constraint (\ref{eq:bd_ris_sc}).
With $\mathbf{\Phi}_l = \mathsf{diag}(\phi_{l,1},\ldots,\phi_{l,M})$, we can rewrite equation (\ref{eq:received_signal}) as 
\begin{equation}
    y_k = \bm{\phi}_l^T\mathbf{Q}_k^H\mathbf{x} + n_k, 
    \label{eq:received_signal1}
\end{equation}
where $\mathbf{Q}_k = \mathbf{G}^H\mathsf{diag}(\mathbf{h}_k)\in\mathbb{C}^{N\times M}$, $\forall k\in\mathcal{K}$ denotes the cascaded channel between the transmitter and user $k$, $\bm{\phi}_l = [\phi_{l,1},\ldots,\phi_{l,M}]^T\in\mathbb{C}^M$, $\forall l\in\mathcal{L}$.

\subsection{Rate-Splitting}
\label{subsec:RSMA}
\begin{figure}
    \centering
    \includegraphics[width=0.48\textwidth]{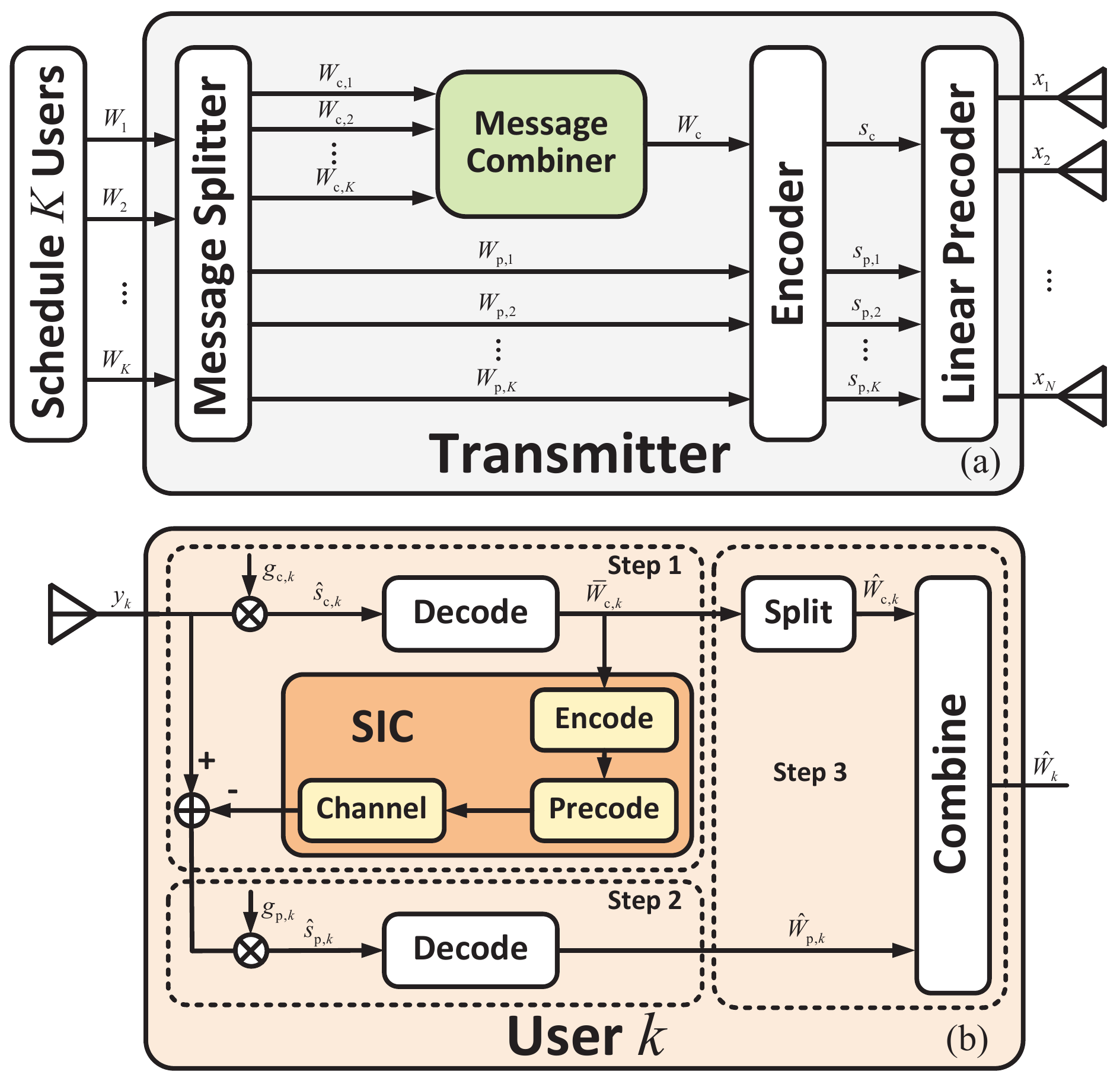}
    \caption{(a) Transmitter and (b) receiver architectures with 1-layer RSMA.}\vspace{-0.2 cm}
    \label{fig:RSMA}
\end{figure}

In this work, we adopt an 1-layer RSMA architecture at the transmitter. 
Fig. \ref{fig:RSMA} illustrates the transceiver architectures with 1-layer RSMA, where the splitting, combining, encoding, and precoding of transmit signals at the transmitter and the decoding, splitting, and combining of received signal at the user side will be described in detail.

\textit{Transmitter:} The message $W_k$ intended to user $k$, $\forall k\in\mathcal{K}$ is split into a common part $W_{\mathrm{c},k}$ and a private part $W_{\mathrm{p},k}$. 
As shown in Fig. \ref{fig:RSMA}(a), the common parts $W_{\mathrm{c},k}$ from all $K$ users are merged into a common message $W_{\mathrm{c}}$ and further encoded into the common stream $s_{\mathrm{c}}$, which will be decoded by all users, by a common codebook. 
$K$ private parts are encoded into private streams $s_{\mathrm{p},1},\ldots,s_{\mathrm{p},K}$ and individually decoded by different users. 
The common stream and $K$ private streams, $\mathbf{s} = [s_{\mathrm{c}},s_{\mathrm{p},1},\ldots,s_{\mathrm{p},K}]^T\in\mathbb{C}^{K+1}$, are linearly precoded by $\mathbf{P} = [\mathbf{p}_{\mathrm{c}},\mathbf{p}_{\mathrm{p},1},\ldots,\mathbf{p}_{\mathrm{p},K}]\in\mathbb{C}^{N\times(K+1)}$, where $\mathbf{p}_{\mathrm{c}}\in\mathbb{C}^N$ is the common precoder and $\mathbf{p}_{\mathrm{p},k}\in\mathbb{C}^N$ is the private precoder for user $k$.
The resulting transmit signal is 
\begin{equation}
    \mathbf{x} = \mathbf{P}\mathbf{s} = \mathbf{p}_{\mathrm{c}}s_\mathrm{c} + \sum\nolimits_{k}\mathbf{p}_{\mathrm{p},k}s_{\mathrm{p},k}.
    \label{eq:transmit_signal}
\end{equation}
Assuming $\mathbb{E}\{\mathbf{s}\mathbf{s}^H\} = \mathbf{I}_{K+1}$, the transmit power constraint  becomes $\mathbb{E}\{\mathbf{x}^H\mathbf{x}\} = \mathsf{Tr}(\mathbf{P}\mathbf{P}^H) = \|\mathbf{P}\|_F^2 \le P$.

\textit{Receiver:} Substituting (\ref{eq:transmit_signal}) into (\ref{eq:received_signal1}), we obtain the received signal at user $k$ covered by sector $l$, $\forall k\in\mathcal{K}_l$,$\forall l\in\mathcal{L}$ as
\begin{equation}
    \begin{aligned}
        y_k =\bm{\phi}_l^T\mathbf{Q}_k^H\mathbf{p}_\mathrm{c}s_\mathrm{c} + \bm{\phi}_l^T\mathbf{Q}_k^H\sum\nolimits_{k'}\mathbf{p}_{\mathrm{p},k'}s_{\mathrm{p},k'}+ n_k.
    \end{aligned}
\end{equation}
Therefore, the average received power of user $k$ covered by sector $l$ for given CSI is given by
\begin{equation}\label{eq:received_power}
    \begin{aligned}
        \tau_{\mathrm{c},k} =& \mathbb{E}\{|y_k|^2\} = |\bm{\phi}_l^T\mathbf{Q}_k^H\mathbf{p}_{\mathrm{c}}|^2\\
        &+\underbrace{|\bm{\phi}_l^T\mathbf{Q}_k^H\mathbf{p}_{\mathrm{p},k}|^2 + \overbrace{\sum\nolimits_{k'\ne k}|\bm{\phi}_l^T\mathbf{Q}_k^H\mathbf{p}_{\mathrm{p},k'}|^2 + \sigma_k^2}^{= \iota_{\mathrm{p},k}}}_{= \tau_{\mathrm{p},k} = \iota_{\mathrm{c},k}}.
    \end{aligned}
\end{equation}
At the user side, the signal process, taking user $k$ covered by sector $l$, $\forall k\in\mathcal{K}_l$, $\forall l\in\mathcal{L}$ as an example, is summarized as following three steps and also illustrated in Fig. \ref{fig:RSMA}(b).

\textit{Step 1:} User $k$ covered by sector $l$ decodes the equalized common stream $\widehat{s}_{\mathrm{c},k} = g_{\mathrm{c},k}y_k$ with equalizer $g_{\mathrm{c},k}\in\mathbb{C}$ into $\bar{W}_{\mathrm{c},k}$ by regarding the interference induced by all $K$ private streams as noise, i.e., $\tau_{\mathrm{p},k}$ or $\iota_{\mathrm{c},k}$ in (\ref{eq:received_power}). After performing SIC, $\bar{W}_{\mathrm{c},k}$ is subtracted from the received signal. 

\textit{Step 2:} User $k$ decodes the equalized private stream $\widehat{s}_{\mathrm{p},k} = g_{\mathrm{p},k}y_k$ with equalizer $g_{\mathrm{p},k}\in\mathbb{C}$ into $\widehat{W}_{\mathrm{p},k}$ by regarding the interference induced by remaining $K-1$ private streams as noise, i.e., $\iota_{\mathrm{p},k}$ in (\ref{eq:received_power}).  

\textit{Step 3:} User $k$ splits its contribution to the common message, $\widehat{W}_{\mathrm{c},k}$, from the decoded $\bar{W}_{\mathrm{c},k}$, and combines $\widehat{W}_{\mathrm{c},k}$ with $\widehat{W}_{\mathrm{p},k}$ to reconstruct $\widehat{W}_k$.

Based on the above signal process, the instantaneous SINRs of the common stream and private streams at user $k$ covered by sector $l$, $\forall k\in\mathcal{K}_l$, $\forall l\in\mathcal{L}$ are given by
\begin{equation}
    \label{eq:SINR}
    \begin{aligned}
        \gamma_{\mathrm{c},k} = |\bm{\phi}_l^T\mathbf{Q}_k^H\mathbf{p}_{\mathrm{c}}|^2\iota_{\mathrm{c},k}^{-1},~
        \gamma_{\mathrm{p},k} = |\bm{\phi}_l^T\mathbf{Q}_k^H\mathbf{p}_{\mathrm{p},k}|^2\iota_{\mathrm{p},k}^{-1}.
    \end{aligned}
\end{equation} 
Assuming Gaussian inputs, the instantaneous common rates and private rates for user $k$ are 
\begin{equation}
    \label{eq:rates}
    R_{\mathrm{c},k} = \log_2(1 + \gamma_{\mathrm{c},k}),~ R_{\mathrm{p},k} = \log_2(1 + \gamma_{\mathrm{p},k}).
\end{equation} 
To guarantee the common message $W_{\mathrm{c}}$ is successfully decoded by all $K$ users, the common rate should not exceed $\min_{\forall k} R_{\mathrm{c},k}$.
Given the architecture of RSMA as illustrated above, the common rate $\min_{\forall k} R_{\mathrm{c},k}$ is split into $K$ parts, yielding $\min_{\forall k} R_{\mathrm{c},k} = \sum_{k}C_k$. Each part $C_k\in\mathbb{R}^+$ corresponds to the common part of the achievable rate for user $k$. 
Finally, the total rate for user $k$ is a summation of common part and private part, i.e., $R_k = C_k + R_{\mathrm{p},k}$, $\forall k\in\mathcal{K}$.

\subsection{Channel State Information}
\label{subsec:CSI}
Perfect CSI acquisition in RIS aided systems is challenging due to the passive property of RIS. Therefore, in this work, we assume the transmitter has an imperfect instantaneous CSI
$\widehat{\mathbf{Q}} = [\widehat{\mathbf{Q}}_1,\ldots,\widehat{\mathbf{Q}}_K]\in\mathbb{C}^{N\times MK}$. To model the CSI uncertainty, the perfect cascaded channel matrix $\mathbf{Q} = [\mathbf{Q}_1,\ldots,\mathbf{Q}_K]\in\mathbb{C}^{N\times MK}$ is expressed as a summation of the channel estimate and channel estimation errors, that is $\mathbf{Q} = \widehat{\mathbf{Q}} + \widetilde{\mathbf{Q}}$, where the estimation error matrix is defined as $\widetilde{\mathbf{Q}} = [\widetilde{\mathbf{Q}}_1,\ldots,\widetilde{\mathbf{Q}}_K]\in\mathbb{C}^{N\times MK}$.
The channel estimation error could come from the noise and limited uplink training in time division duplex (TDD) settings based on the typical MMSE criterion \cite{swindlehurst2022channel}\footnote{The existing popular and effective channel estimation strategy, where the cascaded channel is estimated by pre-defining the RIS pattern during the uplink training process to minimize MSE, can be easily extended to the multi-sector BD-RIS case with CW-SC architecture. The key and main difference compared with conventional RIS case is the RIS pattern design due to the new constraint of multi-sector BD-RIS. A simple choice is to set identical amplitude for non-zero elements of the BD-RIS matrix, i.e., $|\phi_{l,m}| = \frac{1}{\sqrt{L}}$. Then orthogonal matrices, such as discrete Fourier transform (DFT) matrix and Hadamard matrix, can be adopted to construct the RIS pattern.}.
In this case, the channel estimation error generally follows the CSCG distribution, that is $\mathsf{vec}(\widetilde{\mathbf{Q}}_k)\sim\mathcal{CN}(\mathbf{0},\delta_k^2\mathbf{I}_{MN})$, $\forall k\in\mathcal{K}$, where $\delta_k^2$ measures the estimation inaccuracy of the cascaded channel between the transmitter and user $k$. 

\section{Beamforming Design with Sample \\ Average Approximation}
\label{sec:design}

In this section, we formulate the sum-rate maximization problem for BD-RIS aided RSMA based on imperfect CSI conditions and propose efficient algorithms to jointly design the transmit precoder and BD-RIS. 

\subsection{Problem Formulation}
\label{subsec:prob_form}

Beamforming design for RIS aided systems highly relies on the CSI among the transmitter, RIS, and users. 
When perfect CSI is available at the transmitter, beamforming design can be developed according to perfect channels such that the instantaneous sum-rate, $\sum_kR_k$, is maximized, and thus the long-term ergodic sum-rate given by $\mathbb{E}_\mathbf{Q}\{\sum_{k}R_k\}$. 
However, when the transmitter has only partial instantaneous CSI as illustrated in Section \ref{subsec:CSI}, beamforming design that maximizes instantaneous sum-rate as in perfect CSI case is not possible, which makes it difficult to measure the long-term ergodic sum-rate. 
One robust approach to tackle this issue is to perform the beamforming design based on the fact that the transmitter has access to the average rate, which captures the statistical property of the channel estimation error with a given estimate. The average rate is defined as follows. 

\textit{Definition 1:}
The average rate is a measure of the expected performance over a distribution of the channel estimation error, e.g., the distribution of $\widetilde{\mathbf{Q}}$, with determined estimate $\widehat{\mathbf{Q}}$ \cite{joudeh2016sum}. 
Mathematically, the average common and private rates for user $k$, $\forall k\in\mathcal{K}$, depend on the channel estimate $\widehat{\mathbf{Q}}$, and are, respectively, defined as $\bar{R}_{\mathrm{c},k} = \mathbb{E}_{\mathbf{Q}|\widehat{\mathbf{Q}}}\{R_{\mathrm{c},k}|\widehat{\mathbf{Q}}\}$ and $\bar{R}_{\mathrm{p},k} = \mathbb{E}_{\mathbf{Q}|\widehat{\mathbf{Q}}}\{R_{\mathrm{p},k}|\widehat{\mathbf{Q}}\}$, where $R_{\mathrm{c},k}$ and $R_{\mathrm{p},k}$ are instantaneous rates determined by the perfect channel $\mathbf{Q}$ as illustrated in (\ref{eq:rates}).
\hfill$\square$

With Definition 1, we can connect the ergodic rates $\mathbb{E}_\mathbf{Q}\{R_{\mathrm{c},k}\}$, $\mathbb{E}_\mathbf{Q}\{R_{\mathrm{p},k}\}$ with average rates $\bar{R}_{\mathrm{c},k}$, $\bar{R}_{\mathrm{p},k}$ as 
\begin{subequations}
    \begin{align}
        \mathbb{E}_\mathbf{Q}\{R_{\mathrm{c},k}\} = \mathbb{E}_{\widehat{\mathbf{Q}}}\{\mathbb{E}_{\mathbf{Q}|\widehat{\mathbf{Q}}}\{R_{\mathrm{c},k}|\widehat{\mathbf{Q}}\}\} = \mathbb{E}_{\widehat{\mathbf{Q}}}\{\bar{R}_{\mathrm{c},k}\}, \\
        \mathbb{E}_\mathbf{Q}\{R_{\mathrm{p},k}\} = \mathbb{E}_{\widehat{\mathbf{Q}}}\{\mathbb{E}_{\mathbf{Q}|\widehat{\mathbf{Q}}}\{R_{\mathrm{p},k}|\widehat{\mathbf{Q}}\}\} = \mathbb{E}_{\widehat{\mathbf{Q}}}\{\bar{R}_{\mathrm{p},k}\},
    \end{align}
\end{subequations}
which indicate that ergodic rates can be characterized by the expectation of average rates over varied channel estimate $\widehat{\mathbf{Q}}$. 
Accordingly, maximizing the ergodic sum-rate $\mathbb{E}_\mathbf{Q}\{\sum_k R_k\}$ with partial CSI is transformed into maximizing the average sum-rate for each $\widehat{\mathbf{Q}}$, given by $\sum_k\bar{C}_k + \sum_k\bar{R}_{\mathrm{p},k}$, where $\bar{C}_k\in\mathbb{R}^+$, $\forall k\in\mathcal{K}$ referring to the common part of the average rate should satisfy $\sum_{k}\bar{C}_k = \min_{\forall k}\bar{R}_{\mathrm{c},k}$.
Therefore, the average sum-rate maximization problem can be formulated as
\begin{subequations}
    \label{eq:problem0}
    \begin{align}
        \max_{\{\mathbf{P},\bar{\mathbf{c}}\},\{\bm{\phi}_l,\forall l\}} ~~&\sum\nolimits_{k} \bar{C}_k + \sum\nolimits_k \bar{R}_{\mathrm{p},k}\\
        \label{eq:constraint_phi}
        \text{s.t.} ~~~~~~~&\sum\nolimits_{l}|\phi_{l,m}|^2 = 1, \forall m,\\
        \label{eq:constraint_p}
        &\|\mathbf{P}\|_F^2 \le P,\\
        \label{eq:constraint_c1}
        &\sum\nolimits_{k'} \bar{C}_{k'} \le \bar{R}_{\mathrm{c},k}, \forall k,\\
        \label{eq:constraint_c2}
        &\bar{\mathbf{c}} \succeq \mathbf{0},\\
        \label{eq:constraint_qos}
        &\bar{R}_k \ge R_{\mathrm{th},k}, \forall k,
    \end{align}
\end{subequations}
where $\bm{\phi}_l=[\phi_{l,1},\ldots,\phi_{l,M}]^T\in\mathbb{C}^M$, $\forall l\in\mathcal{L}$, $\bar{\mathbf{c}} = [\bar{C}_1,\ldots,\bar{C}_K]^T\in\mathbb{R}^K$, and $R_{\mathrm{th},k}\in\mathbb{R}^+$ is the threshold rate for user $k$, $\forall k\in\mathcal{K}$, to guarantee the QoS.
Problem (\ref{eq:problem0}) is a challenging stochastic optimization different from existing optimization problems due to 1) the new constraint (\ref{eq:constraint_phi}), 2) the constraint of common rates (\ref{eq:constraint_c1}), and 3) the QoS constraint (\ref{eq:constraint_c2}), where the BD-RIS matrix and the precoder are always coupled.
One reliable solution is to 1) transform problem (\ref{eq:problem0}) into a deterministic one based on SAA \cite{shapiro2021lectures}, and to 2) deal with the deterministic problem by well-known WMMSE-rate relationship \cite{christensen2008weighted}. Details of the two-step transformation will be illustrated in Section \ref{subsec:prob_transform}.

\subsection{Two-Step Transformation}
\label{subsec:prob_transform}
\subsubsection{SAA} To facilitate SAA, we give the following definition.

\textit{Definition 2:}
Given an estimate $\widehat{\mathbf{Q}}_k$ of user $k$, a sample of i.i.d. channel realizations is constructed based on the set 
$\mathbb{Q}_k = \{\mathbf{Q}^a_k = \widehat{\mathbf{Q}}_k + \widetilde{\mathbf{Q}}_k^a | \widehat{\mathbf{Q}}_k, \forall a\in\mathcal{A} = \{1,\ldots,A\}\}$,
where $\widetilde{\mathbf{Q}}_k^a$ follows the CSCG distribution, i.e., $\mathsf{vec}(\widetilde{\mathbf{Q}}_k^a)\sim\mathcal{CN}(\mathbf{0},\delta_k^2\mathbf{I}_{MN})$, $\forall a\in\mathcal{A}$. Then the SAA of average common and private rates are, respectively, defined as $\widehat{R}_{\mathrm{c},k} = \frac{1}{A}\sum_{a}R_{\mathrm{c},k}^a$ and $\widehat{R}_{\mathrm{p},k} = \frac{1}{A}\sum_{a}R_{\mathrm{p},k}^a$, $\forall k\in\mathcal{K}$\footnote{In the following illustrations, notations with superscript ``$(\cdot)^a$'' have the same expressions as that without ``$(\cdot)^a$'', except that the term $\mathbf{Q}_k$ is replaced by $\mathbf{Q}_k^a$, $\forall k\in\mathcal{K}_l$, $\forall l\in\mathcal{L}$, $\forall a\in\mathcal{A}$.}.
\hfill$\square$

The SAA of average rate in Definition 2 is a tight approximation of the average rate with sufficient large number of samples\footnote{In this paper, we fix $A=50$ to save on the computational time without sacrificing performance based on simulations using the proposed algorithm and the results in \cite{joudeh2016sum}.}, i.e., $\lim_{A\rightarrow\infty}\widehat{R}_{\mathrm{c},k} = \bar{R}_{\mathrm{c},k}$, $\lim_{A\rightarrow\infty}\widehat{R}_{\mathrm{p},k} = \bar{R}_{\mathrm{p},k}$, $\forall k\in\mathcal{K}$, which motivates us to approximate the stochastic problem (\ref{eq:problem0}) into a deterministic problem
\begin{equation}
    \label{eq:problem1}
    \begin{aligned}
        \max_{\{\mathbf{P},\bar{\mathbf{c}}\},\{\bm{\phi}_l,\forall l\}} ~~&\sum\nolimits_{k}\bar{C}_k + \sum\nolimits_{k}\widehat{R}_{\mathrm{p},k}\\
        \text{s.t.} ~~~~~~~&\text{(\ref{eq:constraint_phi}), (\ref{eq:constraint_p}), (\ref{eq:constraint_c2})}, \\
        &\sum\nolimits_{k'} \bar{C}_{k'} \le \widehat{R}_{\mathrm{c},k}, \forall k,\\
        &\bar{C}_k + \widehat{R}_{\mathrm{p},k} \ge R_{\mathrm{th},k}, \forall k.
    \end{aligned}
\end{equation}
Problem (\ref{eq:problem1}) can be further transformed into a more tractable form based on the following step. 

\subsubsection{WMMSE-Rate Relationship}
Let $\widehat{s}_{\mathrm{c},k}^a = g_{\mathrm{c},k}^ay_k^a$ be the estimate of $s_{\mathrm{c}}$ with equalizer $g_{\mathrm{c},k}^a$ from user $k$ covered by sector $l$, $\forall k\in\mathcal{K}_l$, $\forall l\in\mathcal{L}$, $\forall a\in\mathcal{A}$. 
The estimate of $s_{\mathrm{p},k}$ with equalizer $g_{\mathrm{p},k}^a$ after removing the common stream is given by $\widehat{s}_{\mathrm{p},k}^a = g_{\mathrm{p},k}^a(y_k^a - \bm{\phi}_{l}^T(\mathbf{Q}_k^a)^H\mathbf{p}_{\mathrm{c}}s_{\mathrm{c}})$, $\forall k\in\mathcal{K}_l$, $\forall l\in\mathcal{L}$, $\forall a\in\mathcal{A}$.
Define MSE functions regarding the common and private streams for user $k$ covered by sector $l$ at channel sample $a$ as $\epsilon_{\mathrm{c},k}^a = \mathbb{E}\{|\widehat{s}_{\mathrm{c},k}^a - s_{\mathrm{c}}|^2\} = |g_{\mathrm{c},k}^a|^2\tau_{\mathrm{c},k}^a -2\Re\{g_{\mathrm{c},k}^a\bm{\phi}_l^T(\mathbf{Q}_k^a)^H\mathbf{p}_{\mathrm{c}}\} + 1$ and $\epsilon_{\mathrm{p},k}^a = \mathbb{E}\{|\widehat{s}_{\mathrm{p},k}^a - s_{\mathrm{p},k}|^2\} = |g_{\mathrm{p},k}^a|^2\tau_{\mathrm{p},k}^a -2\Re\{g_{\mathrm{p},k}^a\bm{\phi}_l^T(\mathbf{Q}_k^a)^H\mathbf{p}_{\mathrm{p},k}\} + 1$.
Then optimal unconstrained equalizers for the common stream and private streams which achieve minimum MSEs are given by 
\begin{equation}
    \label{eq:opt_equalizer}
        (g_{\mathrm{c},k}^a)^\star = \frac{\mathbf{p}_{\mathrm{c}}^H\mathbf{Q}_k^a\bm{\phi}_l^*}{\tau_{\mathrm{c},k}^a}, ~
        (g_{\mathrm{p},k}^a)^\star = \frac{\mathbf{p}_{\mathrm{p},k}^H\mathbf{Q}_k^a\bm{\phi}_l^*}{\tau_{\mathrm{p},k}^a}, 
\end{equation} 
and the corresponding minimum MSEs are $(\epsilon_{o,k}^a)^\star = \min_{g_{o,k}^a} \epsilon_{o,k}^a = (\tau_{o,k}^a)^{-1}\iota_{o,k}^a = (1 + \gamma_{o,k}^a)^{-1}$, $\forall o\in\{\mathrm{c,p}\}$, $\forall k\in\mathcal{K}$, $\forall a\in\mathcal{A}$.
Introducing weights $\lambda_{\mathrm{c},k}^a\in\mathbb{R}^+$ and $\lambda_{\mathrm{p},k}^a\in\mathbb{R}^+$ associating common and private MSEs for user $k$, we can establish the WMMSE-rate relationship $R_{o,k}^a = \max_{\lambda_{o,k}^a} \log_2\lambda_{o,k}^a - \lambda_{o,k}^a(\epsilon_{o,k}^a)^\star + 1$, $\forall o\in\{\mathrm{c,p}\}$, $\forall k\in\mathcal{K}$, $\forall a\in\mathcal{A}$,
where the maximum of the right-hand problem is achieved when weights $\lambda_{\mathrm{c},k}$ and $\lambda_{\mathrm{p},k}$ satisfy
\begin{equation}
    \label{eq:opt_weight}
        (\lambda_{\mathrm{c},k}^a)^\star = 1 + \gamma_{\mathrm{c},k}^a, ~(\lambda_{\mathrm{p},k}^a)^\star = 1 + \gamma_{\mathrm{p},k}^a.
\end{equation}

Based on the above derivations, we reformulate problem (\ref{eq:problem1}) as a three-block optimization 
\begin{subequations}
    \label{eq:problem2}
    \begin{align}
        \non
        \max_{\substack{\{\bm{\lambda}_o^a, \mathbf{g}_o^a,\forall a,\forall o\}\\ \{\mathbf{P},\bar{\mathbf{c}}\}, \{\bm{\phi}_l,\forall l\}}} & \sum\nolimits_{k}\left(\bar{C}_k
        + \frac{1}{A}\sum\nolimits_{a}(\log_2\lambda_{\mathrm{p},k}^a-\lambda_{\mathrm{p},k}^a\epsilon_{\mathrm{p},k}^a)\right)\\
        \non
        \text{s.t.} ~~~~~&\text{(\ref{eq:constraint_phi}), (\ref{eq:constraint_p}), (\ref{eq:constraint_c2})},\\
        \non
        &\frac{1}{A}\sum\nolimits_{a}(\log_2\lambda_{\mathrm{c},k}^a-\lambda_{\mathrm{c},k}^a\epsilon_{\mathrm{c},k}^a+1)\\
        \label{eq:constraint_c1_modified}
        &~~~~~\ge \sum\nolimits_{k'} \bar{C}_{k'}, \forall k,\\
        \non
        &\bar{C}_k + \frac{1}{A}\sum\nolimits_{a}(\log_2\lambda_{\mathrm{p},k}^a-\lambda_{\mathrm{p},k}^a\epsilon_{\mathrm{p},k}^a+1)\\
        \label{eq:constraint_qos_modified}
        &~~~~~\ge R_{\mathrm{th},k}, \forall k,
    \end{align}
\end{subequations}
where $\bm{\lambda}_o^a = [\lambda_{o,1}^a,\ldots,\lambda_{o,K}^a]^T$ and $\mathbf{g}_o^a = [g_{o,1}^a,\ldots,g_{o,K}^a]^T$, $\forall o\in\{\mathrm{c,p}\}$, $\forall a\in\mathcal{A}$. 
Although problem (\ref{eq:problem2}) is not convex in a joint design manner, the sub-problem regarding one block with fixing other blocks is either convex, such as the design of blocks $\{\bm{\lambda}_o^a, \mathbf{g}_o^a,\forall o, \forall a\}$ and $\{\mathbf{P},\bar{\mathbf{c}}\}$, or more tractable, such as the design of block $\{\bm{\phi}_l,\forall l\in\mathcal{L}\}$, than directly dealing with the original objective. Specifically, the unconstrained solution to block $\{\bm{\lambda}_o^a, \mathbf{g}_o^a, \forall o, \forall a\}$ with fixed two other blocks have been given in equations (\ref{eq:opt_equalizer}) and (\ref{eq:opt_weight}), respectively. Solutions to blocks $\{\mathbf{P},\bar{\mathbf{c}}\}$ and $\{\bm{\phi}_l,\forall l\}$ will be respectively given in Sections \ref{subsec:precoder_design} and \ref{subsec:BD-RIS_design}.

\subsection{Solution to Block $\{\mathbf{P},\bar{\mathbf{c}}\}$}
\label{subsec:precoder_design}
With fixed equalizers, weights, and BD-RIS matrix, the sub-problem regarding block $\{\mathbf{P},\bar{\mathbf{c}}\}$ is given by 
\begin{equation}
    \label{eq:sub_p0}
    \begin{aligned}
        \min_{\{\mathbf{P},\bar{\mathbf{c}}\}} ~~&\frac{1}{A}\sum\nolimits_{k}\sum\nolimits_{a} \lambda_{\mathrm{p},k}^a\epsilon_{\mathrm{p},k}^a - \sum\nolimits_{k}\bar{C}_k\\
        \text{s.t.}~~~&\text{(\ref{eq:constraint_p}), (\ref{eq:constraint_c2}), (\ref{eq:constraint_c1_modified}), (\ref{eq:constraint_qos_modified})}.
    \end{aligned}
\end{equation}
With the following definitions 
\begin{subequations}
    \begin{align}
        \bm{\psi}_{\mathrm{c},k}^a &= \lambda_{\mathrm{c},k}^a(g_{\mathrm{c},k}^a)^*\mathbf{Q}_k^a\bm{\phi}_l^*, ~\bm{\psi}_{\mathrm{p},k}^a = \lambda_{\mathrm{p},k}^a(g_{\mathrm{p},k}^a)^*\mathbf{Q}_k^a\bm{\phi}_l^*,\\
        % \bm{\psi}_{\mathrm{c},k} &= \frac{1}{A}\sum\nolimits_{a} \bm{\psi}_{\mathrm{c},k}^a, ~\bm{\psi}_{\mathrm{p},k} = \frac{1}{A}\sum\nolimits_{a} \bm{\psi}_{\mathrm{p},k}^a,\\
        \mathbf{\Psi}_{\mathrm{c},k}^a &= \frac{\bm{\psi}_{\mathrm{c},k}^a(\bm{\psi}_{\mathrm{c},k}^a)^H}{\lambda_{\mathrm{c},k}^a}, ~\mathbf{\Psi}_{\mathrm{p},k}^a = \frac{\bm{\psi}_{\mathrm{p},k}^a(\bm{\psi}_{\mathrm{p},k}^a)^H}{\lambda_{\mathrm{p},k}^a},\\
        % \mathbf{\Psi}_{\mathrm{c},k} &= \frac{1}{A}\sum\nolimits_{a} \mathbf{\Psi}_{\mathrm{c},k}^a, ~\mathbf{\Psi}_{\mathrm{p},k} = \frac{1}{A}\sum\nolimits_{a} \mathbf{\Psi}_{\mathrm{p},k}^a,\\
        \xi_{\mathrm{c},k}^a &= \log_2\lambda_{\mathrm{c},k}^a - \lambda_{\mathrm{c},k}^a|g_{\mathrm{c},k}^a|^2\sigma_k^2 - \lambda_{\mathrm{c},k}^a + 1,\\
        \xi_{\mathrm{p},k}^a &= \log_2\lambda_{\mathrm{p},k}^a - \lambda_{\mathrm{p},k}^a|g_{\mathrm{p},k}^a|^2\sigma_k^2 - \lambda_{\mathrm{p},k}^a + 1,
        % \xi_{\mathrm{c},k} &= \frac{1}{A}\sum\nolimits_{a} (\log_2\lambda_{\mathrm{c},k}^a - \lambda_{\mathrm{c},k}^a|g_{\mathrm{c},k}^a|^2\sigma_k^2 - \lambda_{\mathrm{c},k}^a) + 1,\\
        % \non
        % \xi_{\mathrm{p},k} &= \frac{1}{A}\sum\nolimits_{a} (\log_2\lambda_{\mathrm{p},k}^a - \lambda_{\mathrm{p},k}^a|g_{\mathrm{p},k}^a|^2\sigma_k^2 - \lambda_{\mathrm{p},k}^a)\\
        % &~~~~+ 1-R_{\mathrm{th},k},\\
        % \grave{\mathbf{c}} &= [\grave{C}_1,\ldots,\grave{C}_K]^T = -\bar{\mathbf{c}},
    \end{align}
\end{subequations}
and their averages over $A$ samples, i.e., $\bm{\psi}_{\mathrm{c},k} = \frac{1}{A}\sum\nolimits_{a} \bm{\psi}_{\mathrm{c},k}^a$, $\bm{\psi}_{\mathrm{p},k} = \frac{1}{A}\sum\nolimits_{a} \bm{\psi}_{\mathrm{p},k}^a$, $\mathbf{\Psi}_{\mathrm{c},k} = \frac{1}{A}\sum\nolimits_{a} \mathbf{\Psi}_{\mathrm{c},k}^a$, $\mathbf{\Psi}_{\mathrm{p},k} = \frac{1}{A}\sum\nolimits_{a} \mathbf{\Psi}_{\mathrm{p},k}^a$, $\xi_{\mathrm{c},k} = \frac{1}{A}\sum\nolimits_{a} \xi_{\mathrm{c},k}^a$, $\xi_{\mathrm{p},k} = \frac{1}{A}\sum\nolimits_{a} \xi_{\mathrm{p},k}$, $\forall k\in\mathcal{K}$,
problem (\ref{eq:sub_p0}) can be rewritten as 
\begin{subequations}
    \label{eq:sub_p1}
    \begin{align}
        \non
        \min_{\grave{\mathbf{c}},\mathbf{p}_\mathrm{c},\{\mathbf{p}_{\mathrm{p},k},\forall k\}} &\sum\nolimits_{k}\Big(\grave{C}_k+\sum\nolimits_{k'}\mathbf{p}_{\mathrm{p},k'}^H\mathbf{\Psi}_{\mathrm{p},k}\mathbf{p}_{\mathrm{p},k'}\\
         &~~~~- 2\Re\{\bm{\psi}_{\mathrm{p},k}^H\mathbf{p}_{\mathrm{p},k}\}\Big)\\
        \text{s.t.}~~~~&\|\mathbf{p}_{\mathrm{c}}\|_2^2 + \sum\nolimits_{k}\|\mathbf{p}_{\mathrm{p},k}\|_2^2 \le P,\\
        \non
        &\mathbf{p}_{\mathrm{c}}^H\mathbf{\Psi}_{\mathrm{c},k}\mathbf{p}_{\mathrm{c}} + \sum\nolimits_{k'}\mathbf{p}_{\mathrm{p},k'}^H\mathbf{\Psi}_{\mathrm{c},k}\mathbf{p}_{\mathrm{p},k'}\\
        &~-2\Re\{\bm{\psi}_{\mathrm{c},k}^H\mathbf{p}_{\mathrm{c}}\} \le \sum\nolimits_{k'}\grave{C}_k + \xi_{\mathrm{c},k},\forall k,\\
        &\grave{\mathbf{c}} \preceq \mathbf{0},\\
        \non
        &\sum\nolimits_{k'}\mathbf{p}_{\mathrm{p},k'}^H\mathbf{\Psi}_{\mathrm{p},k}\mathbf{p}_{\mathrm{p},k'}- 2\Re\{\bm{\psi}_{\mathrm{p},k}^H\mathbf{p}_{\mathrm{p},k}\}\\
        &~ + \grave{C}_k\le \xi_{\mathrm{p},k} - R_{\mathrm{th},k}, \forall k,
    \end{align}
\end{subequations}
where $\grave{\mathbf{c}} = [\grave{C}_1,\ldots,\grave{C}_K]^T = -\bar{\mathbf{c}}$. Problem (\ref{eq:sub_p1}) is a convex second-order cone program (SOCP) and can be solved by interior-point methods with computational complexity bounded by $\mathcal{O}((NK+N+K)^{3.5})$ \cite{boyd2004convex}.

\subsection{Solution to Block $\{\bm{\phi}_l,\forall l\}$}
\label{subsec:BD-RIS_design}
Recall that (\ref{eq:constraint_qos_modified}) in problem (\ref{eq:problem2}) is the QoS constraint, which can be guaranteed in the design of block $\{\mathbf{P},\bar{\mathbf{c}}\}$. To simplify the design of block $\{\bm{\phi}_l,\forall l\}$, we temporarily ignore constraint (\ref{eq:constraint_qos_modified}). Then, without the QoS constraint (\ref{eq:constraint_qos_modified}), we can remove the term $\bar{\mathbf{c}}$ by putting constraint (\ref{eq:constraint_c1_modified}) into the objective. In this case, the original problem (\ref{eq:problem2}) can be simplified as the following form
\begin{subequations}
    \label{eq:problem3}
    \begin{align}
        \non
        \max_{\substack{\{\bm{\lambda}_o^a, \mathbf{g}_o^a, \forall o, \forall a\}\\ \mathbf{P}, \{\bm{\phi}_l,\forall l\}}} & \left\{\min\nolimits_{\forall k} \frac{1}{A}\sum\nolimits_{a}(\log_2\lambda_{\mathrm{c},k}^a-\lambda_{\mathrm{c},k}^a\epsilon_{\mathrm{c},k}^a+1)\right\}\\
        &+\frac{1}{A}\sum\nolimits_{k}\sum\nolimits_{a}(\log_2\lambda_{\mathrm{p},k}^a-\lambda_{\mathrm{p},k}^a\epsilon_{\mathrm{p},k}^a)\\
        \text{s.t.} ~~~~~&\text{(\ref{eq:constraint_phi}), (\ref{eq:constraint_p}), (\ref{eq:constraint_c2})}.
    \end{align}
\end{subequations}
With given equalizers, weights, and the transmit precoder, the sub-problem regarding the multi-sector BD-RIS is 
\begin{equation}
    \label{eq:sub_phi0}
    \begin{aligned}
        \min_{\{\bm{\phi}_l,\forall l\}} ~&\left\{\max\nolimits_{\forall k} \frac{1}{A}\sum\nolimits_{a}(\lambda_{\mathrm{c},k}^a\epsilon_{\mathrm{c},k}^a-\log_2\lambda_{\mathrm{c},k}^a-1)\right\} \\ &+ \frac{1}{A}\sum\nolimits_{a}\sum\nolimits_{k} \lambda_{\mathrm{p},k}^a\epsilon_{\mathrm{p},k}^a\\
        \text{s.t.}~~~&\sum\nolimits_{l}|\phi_{l,m}|^2 = 1, \forall m.
    \end{aligned}
\end{equation}
With the following definitions
\begin{subequations}
    \begin{align}
        \mathbf{v}_{\mathrm{c},k}^a =& (\mathbf{Q}_k^a)^T\mathbf{p}_{\mathrm{c}}^*, ~~
        \mathbf{v}_{\mathrm{p},k,k'}^a = (\mathbf{Q}_k^a)^T\mathbf{p}_{\mathrm{p},k'}^*, \\
        \non
        \mathbf{X}_{\mathrm{c},k}^a =& \lambda_{\mathrm{c},k}^a|g_{\mathrm{c},k}^a|^2\mathbf{v}_{\mathrm{c},k}^a(\mathbf{v}_{\mathrm{c},k}^a)^H \\
        &+ \lambda_{\mathrm{c},k}^a|g_{\mathrm{c},k}^a|^2\sum\nolimits_{k'}\mathbf{v}_{\mathrm{p},k,k'}^a(\mathbf{v}_{\mathrm{p},k,k'}^a)^H,\\
        \mathbf{X}_{\mathrm{p},k}^a =& \lambda_{\mathrm{p},k}^a|g_{\mathrm{p},k}^a|^2\sum\nolimits_{k'}\mathbf{v}_{\mathrm{p},k,k'}^a(\mathbf{v}_{\mathrm{p},k,k'}^a)^H,\\ 
        % \mathbf{X}_{\mathrm{c},k} =& \frac{1}{A}\sum_{a\in\mathcal{A}}\mathbf{X}_{\mathrm{c},k}^a, ~ \mathbf{X}_{\mathrm{p},k} = \frac{1}{A}\sum_{a\in\mathcal{A}}\mathbf{X}_{\mathrm{p},k}^a,\\
        \mathbf{x}_{\mathrm{c},k}^a =& \lambda_{\mathrm{c},k}^a(g_{\mathrm{c},k}^a)^*\mathbf{v}_{\mathrm{c},k}^a,~
        \mathbf{x}_{\mathrm{p},k}^a = \lambda_{\mathrm{p},k}^a(g_{\mathrm{p},k}^a)^*\mathbf{v}_{\mathrm{p},k,k}^a, 
        % \mathbf{x}_{\mathrm{c},k} =& \frac{1}{A}\sum_{a\in\mathcal{A}}\mathbf{x}_{\mathrm{c},k}^a, ~ \mathbf{x}_{\mathrm{p},k} = \frac{1}{A}\sum_{a\in\mathcal{A}}\mathbf{x}_{\mathrm{p},k}^a, \\
        % \bar{\xi}_{\mathrm{p},k} =& \xi_{\mathrm{p},k} - \grave{C}_k,
    \end{align}
\end{subequations}
and their averages over $A$ samples, i.e., $\mathbf{X}_{\mathrm{c},k} = \frac{1}{A}\sum_{a}\mathbf{X}_{\mathrm{c},k}^a$, $\mathbf{X}_{\mathrm{p},k} = \frac{1}{A}\sum_{a}\mathbf{X}_{\mathrm{p},k}^a$, $\mathbf{x}_{\mathrm{c},k} = \frac{1}{A}\sum_{a}\mathbf{x}_{\mathrm{c},k}^a$, $\mathbf{x}_{\mathrm{p},k} = \frac{1}{A}\sum_{a}\mathbf{x}_{\mathrm{p},k}^a$, $\forall k\in\mathcal{K}$, 
problem (\ref{eq:sub_phi0}) can be rearranged as 
\begin{equation}
    \label{eq:sub_phi2}
    \begin{aligned}
        \min_{\{\bm{\phi}_l,\forall l\}}~~&\left\{\max\nolimits_{\forall k}\left(\bm{\phi}_l^H\mathbf{X}_{\mathrm{c},k}\bm{\phi}_l - 2\Re\{\bm{\phi}_l^H\mathbf{x}_{\mathrm{c},k}\} - \xi_{\mathrm{c},k}\right)\right\}\\
         &+ \sum\nolimits_{l}\left(\bm{\phi}_l^H\bar{\mathbf{X}}_{\mathrm{p},l}\bm{\phi}_l - 2\Re\{\bm{\phi}_l^H\bar{\mathbf{x}}_{\mathrm{p},l}\}\right)\\
        \text{s.t.}~~~~~&\sum\nolimits_{l}|\phi_{l,m}|^2 = 1, \forall m,
    \end{aligned}
\end{equation}
where $\bar{\mathbf{X}}_{\mathrm{p},l} = \sum_{k'\in\mathcal{K}_l}\mathbf{X}_{\mathrm{p},k'}$ and $\bar{\mathbf{x}}_{\mathrm{p},l} = \sum_{k'\in\mathcal{K}_l}\mathbf{x}_{\mathrm{p},k'}$, $\forall l\in\mathcal{L}$.
We observe from problem (\ref{eq:sub_phi2}) that the objective regarding each cell of the multi-sector BD-RIS is coupled with each other, while the constraints for different cells are independent of each other. This fact motivates us to successively design each cell while fixing the rest cells. Specifically, the design of one cell with fixed others will be given as follows.

We first formulate the sub-problem regarding cell $m$, $\forall m\in\mathcal{M}$, with fixed other cells as 
\begin{equation}
    \label{eq:sub_phi_m}
    \begin{aligned}
        \min_{\bar{\bm{\phi}}_m}~~&\left\{\max\nolimits_{\forall k} \underbrace{\nu_{\mathrm{c},k,m}|\phi_{l,m}|^2 - 2\Re\{\phi_{l,m}^*\chi_{\mathrm{c},k,m}\} - \xi_{k,m}}_{=f_{\mathrm{c},k}(\phi_{l,m})}\right\}\\
        & + \sum\nolimits_{l} \underbrace{\nu_{\mathrm{p},l,m}|\phi_{l,m}|^2 - 2\Re\{\phi_{l,m}^*\chi_{\mathrm{p},l,m}\}}_{=f_{\mathrm{p},l}(\phi_{l,m})}\\
        \text{s.t.}~~~&\|\bar{\bm{\phi}}_m\|_2^2 = 1,   
    \end{aligned}
\end{equation}
where the fresh notations are respectively defined as 
\begin{subequations}
    \label{eq:aux}
    \begin{align}
        \nu_{\mathrm{c},k,m} &= [\mathbf{X}_{\mathrm{c},k}]_{m,m}, ~ \nu_{\mathrm{p},l,m} = [\bar{\mathbf{X}}_{\mathrm{p},l}]_{m,m},\\
        \chi_{\mathrm{c},k,m} &= [\mathbf{x}_{\mathrm{c},k}]_m - \sum\nolimits_{n\ne m}[\mathbf{X}_{\mathrm{c},k}]_{m,n}\phi_{l,n},\\
        \chi_{\mathrm{p},l,m} &=  [\bar{\mathbf{x}}_{\mathrm{p},l}]_m - \sum\nolimits_{n\ne m}[\bar{\mathbf{X}}_{\mathrm{p},l}]_{m,n}\phi_{l,n},\\
        \non
        \xi_{k,m} &= \xi_{\mathrm{c},k} - \sum\nolimits_{n\ne m}\sum\nolimits_{n'\ne m}[\mathbf{X}_{\mathrm{c},k}]_{n,n'}\phi_{l,n}^*\phi_{l,n'}\\
        &~~~ + 2\Re\left\{\sum\nolimits_{n\ne m}\phi_{l,n}^*[\mathbf{x}_{\mathrm{c},k}]_n\right\},
    \end{align}
\end{subequations}
and $\bar{\bm{\phi}}_m = [\phi_{1,m},\ldots,\phi_{L,m}]$, $\forall m\in\mathcal{M}$, $\forall k\in\mathcal{K}_l$, $\forall l\in\mathcal{L}$.
Problem (\ref{eq:sub_phi_m}) is a non-smooth optimization with a non-convex constraint. 
To solve this problem, we 1) introduce the well-known LogSumExp (LSE) function to transform the non-smooth part of the objective function into a smooth form, and 2) adopt the manifold theory \cite{boumal2014optimization} to transform the constrained problem on the Euclidean space into an unconstrained optimization on the manifold space. Each of the two steps is described as follows.

\textit{Step 1: LSE Approximation.} To facilitate the LSE approximation, we give the following definition. 

\textit{Definition 3:}
Given a set $\mathbb{F} = \{f_{\mathrm{c},k}\in\mathbb{R} | \forall k\in\mathcal{K}\}$, an LSE function is defined as the logarithm of the sum of the exponentials of $f_{\mathrm{c},k}$, $\forall k$, i.e., $\log\sum_{k}\exp(f_{\mathrm{c},k})$, which is a smooth approximation of the maximum $\max_{\forall k}f_{\mathrm{c},k}$ with a lower bound $\max_{\forall k}f_{\mathrm{c},k}$ and an upper bound $\max_{\forall k}f_{\mathrm{c},k} + \log|\mathcal{K}|$, i.e., $\max_{\forall k}f_{\mathrm{c},k}\le \log\sum_{k}\exp(f_{\mathrm{c},k})\le \max_{\forall k}f_{\mathrm{c},k} + \log|\mathcal{K}|$.  

\textit{Proof:}
Let $f_\mathrm{c}^\mathrm{max} = \max_{\forall k} f_{\mathrm{c},k}$. Then we have the inequality $\exp(f_\mathrm{c}^\mathrm{max})\le\sum_k\exp(f_{\mathrm{c},k})\le|\mathcal{K}|\exp(f_\mathrm{c}^\mathrm{max})$. Performing the logarithm to the inequality gives the result $\max_{\forall k}f_{\mathrm{c},k}\le \log\sum_{k}\exp(f_{\mathrm{c},k})\le \max_{\forall k}f_{\mathrm{c},k} + \log|\mathcal{K}|$.
\hfill$\square$

With Definition 3, we have the following corollary to obtain an approximation to $\max_{\forall k} f_{\mathrm{c},k}(\phi_{l,m})$ in problem (\ref{eq:sub_phi_m}).

\textit{Corollary 1:}
$\max_{\forall k} f_{\mathrm{c},k}(\phi_{l,m})$ has the following bounds 
\begin{equation}
    \label{eq:LSE}
    \begin{aligned}
        \max\nolimits_{\forall k} f_{\mathrm{c},k}(\phi_{l,m}) &\overset{\text{(a)}}{\le} \underbrace{\varepsilon\log\sum\nolimits_{k}\exp\left(\frac{f_{\mathrm{c},k}(\phi_{l,m})}{\varepsilon}\right)}_{=\bar{f}(\bar{\bm{\phi}}_m)}\\
        &\overset{\text{(b)}}{\le} \max\nolimits_{\forall k} f_{\mathrm{c},k}(\phi_{l,m}) + \varepsilon\log|\mathcal{K}|,
    \end{aligned}
\end{equation}
with $\varepsilon>0$, where (a) achieves its equality when $|\mathcal{K}| = 1$ and (b) achieves its equality when $f_{\mathrm{c},1}(\phi_{l,m}) =\ldots= f_{\mathrm{c},K}(\phi_{l,m})$.

\textit{Proof:}
Based on Definition 1, by replacing each $f_{\mathrm{c},k}$ with $\frac{f_{\mathrm{c},k}}{\varepsilon}$, $\forall k\in\mathcal{K}$, we have the inequality $\max_{\forall k} \frac{f_{\mathrm{c},k}}{\varepsilon} \le \log\sum\nolimits_{k}\exp(\frac{f_{\mathrm{c},k}}{\varepsilon})\le \max\nolimits_{\forall k} \frac{f_{\mathrm{c},k}}{\varepsilon} + \log|\mathcal{K}|$. Multiplying by $\varepsilon$ gives the inequality (\ref{eq:LSE}).
\hfill$\square$

\begin{figure}[t]
    \centering
    \includegraphics[width = 0.48\textwidth]{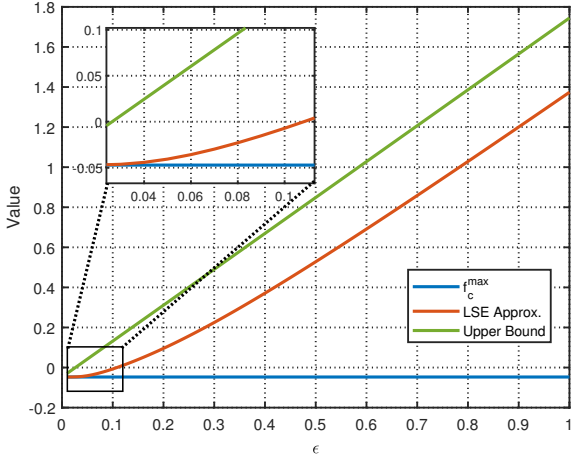}
    \caption{The impact of $\varepsilon$ to the tightness of LSE approximation.}\vspace{-0.2 cm}
    \label{fig:LSE_approx}
\end{figure}

In Corollary 1, $\bar{f}(\bar{\bm{\phi}})$ refers to the LSE function of $\max_{\forall k}f_{\mathrm{c},k}(\phi_{l,m})$ with $\varepsilon$ a scalar to make the bounds tighter. In addition, we can easily deduce that the smaller the $\varepsilon$, the tighter the LSE approximation. To get some insights of how the $\varepsilon$ influences the tightness of the bounds, we compare the values of $f_\mathrm{c}^\mathrm{max} = \max_{\forall k}f_{\mathrm{c},k}(\phi_{l,m})$, its LSE approximation $\bar{f}(\bar{\bm{\phi}}_m)$, and the upper bound $f_\mathrm{c}+\varepsilon\log|\mathcal{K}|$ with different $\varepsilon$ in Fig. \ref{fig:LSE_approx}, which verifies that a smaller $\varepsilon$ leads to a smaller gap between $f_\mathrm{c}^\mathrm{max}$ and the corresponding LSE function.

With Corollary 1, problem (\ref{eq:sub_phi_m}) can be approximated as 
\begin{equation}
    \label{eq:sub_phi_m1}
    \bar{\bm{\phi}}_m^\star = \arg\min_{\|\bar{\bm{\phi}}_m\|_2^2 = 1} ~\underbrace{\sum\nolimits_{l} f_{\mathrm{p},l}(\phi_{l,m}) + \bar{f}(\bar{\bm{\phi}}_m)}_{=f(\bar{\bm{\phi}}_m)}.
\end{equation}
Now problem (\ref{eq:sub_phi_m1}) has a smooth objective function $f(\bar{\bm{\phi}}_m)$ and only one constraint, which can be solved by the manifold method \cite{boumal2014optimization} as illustrated below.

\textit{Step 2: Manifold Construction.} 
We give the following definition to construct the feasible region of problem (\ref{eq:sub_phi_m1}), i.e., $\|\bar{\bm{\phi}}_m\|_2^2 = 1$, as a manifold \cite{boumal2014optimization}, \cite{petersen2006riemannian}. 

\textit{Definition 4:}
An $L$-dimensional complex sphere manifold is defined as $\mathbb{M} = \{\bar{\bm{\phi}}_m\in\mathbb{C}^{1\times L}: \|\bar{\bm{\phi}}_m\|_2 = 1\}$, where $\bar{\bm{\phi}}_m$ is a point on the complex sphere manifold $\mathbb{M}$. 
Given a point $\bar{\bm{\phi}}_m$ on $\mathbb{M}$, a tangent vector refers to one possible moving direction of this point. The set of these tangent vectors regarding point $\bar{\bm{\phi}}_m$ forms the tangent space, which is mathematically defined as $\mathbb{T}_{\bar{\bm{\phi}}_m}\mathbb{M} = \{\mathbf{u}_m\in\mathbb{C}^{1\times L}: \Re\{\mathsf{Tr}(\bar{\bm{\phi}}_m^H\mathbf{u}_m)\} = 0\}$. 
Among all these tangent vectors on $\mathbb{T}_{\bar{\bm{\phi}}_m}\mathbb{M}$, Riemannian gradient is defined as the tangent vector with the steepest-descent direction of the objective function.
\hfill$\square$

With Definition 4, we say $\bar{\bm{\phi}}_m$ is a point on the complex sphere manifold $\mathbb{M}$, yielding problem (\ref{eq:sub_phi_m1}) an unconstrained optimization on $\mathbb{M}$, which can be solved by many classic algorithms with necessary projections. 
The algorithm we use here to solve problem (\ref{eq:sub_phi_m1}) is the Riemannian conjugate gradient (RCG) algorithm \cite{boumal2014optimization}, which is a modified version of the well-known CG algorithm by changing the search space from the Euclidean space to a manifold as detailed below.  

The key of RCG is to determine the descent direction of the objective function. 
In RCG, the Riemannian gradient, which can be obtained from the Euclidean gradient, is required to calculate the descent direction. Therefore, we first calculate the Euclidean gradient of the objective function 
\begin{equation}
    \begin{aligned}
    f(\bar{\bm{\phi}}_m) 
    &= \sum\nolimits_{l} f_{\mathrm{p},l}(\phi_{l,m}) + \bar{f}(\bar{\bm{\phi}}_m)\\
    &= \sum\nolimits_{l} f_{\mathrm{p},l}(\phi_{l,m}) + \varepsilon\log\sum\nolimits_k\exp\left(\frac{f_{\mathrm{c},k}(\phi_{l,m})}{\varepsilon}\right),
    \end{aligned}
\end{equation} 
which is given by
\begin{subequations}
    \begin{align}
        \non
        \triangledown f(\bar{\bm{\phi}}_m) &= \left[\frac{\partial f(\bar{\bm{\phi}}_m)}{\partial \phi_{1,m}}, \ldots, \frac{\partial f(\bar{\bm{\phi}}_m)}{\partial \phi_{L,m}}\right]\\
        \non
        &= \Big[\triangledown f_{\mathrm{p},1}(\phi_{1,m}) + \frac{\partial \bar{f}(\bar{\bm{\phi}}_m)}{\partial \phi_{1,m}},\\
        & ~~~~~\ldots, \triangledown f_{\mathrm{p},L}(\phi_{L,m}) + \frac{\partial \bar{f}(\bar{\bm{\phi}}_m)}{\partial \phi_{L,m}}\Big],\\
        \triangledown f_{\mathrm{p},l}(\phi_{l,m}) &= 2\nu_{\mathrm{p},l,m}\phi_{l,m} - 2\chi_{\mathrm{p},l,m},\\
        \frac{\partial \bar{f}(\bar{\bm{\phi}}_m)}{\partial \phi_{l,m}} &= \frac{\sum_{k\in\mathcal{K}_l}\exp\left(\frac{{f}_{\mathrm{c},k}(\phi_{l,m})}{\varepsilon}\right)\triangledown f_{\mathrm{c},k}(\phi_{l,m})}{\sum_{k\in\mathcal{K}}\exp\left(\frac{{f}_{\mathrm{c},k}(\phi_{l,m})}{\varepsilon}\right)}, \\
        \triangledown f_{\mathrm{c},k}(\phi_{l,m}) &= 2\nu_{\mathrm{c},k,m}\phi_{l,m} - 2\chi_{\mathrm{c},k,m}.
    \end{align}
\end{subequations}
Then, we obtain the Riemannian gradient, which projects the Euclidean gradient onto the tangent space, i.e., $\triangledown_{\mathbb{M}} f(\bar{\bm{\phi}}_m) = \mathsf{Pr}_{\bar{\bm{\phi}}_m}(\triangledown f(\bar{\bm{\phi}}_m)) = \triangledown f(\bar{\bm{\phi}}_m) - \Re\{\mathsf{Tr}(\bar{\bm{\phi}}_m^H\triangledown f(\bar{\bm{\phi}}_m))\bar{\bm{\phi}}_m\}$. 
The descent direction at the $v$-th iteration of RCG is thus
\begin{equation}
    \label{eq:xi}
    \bm{\beta}^v = -\triangledown_{\mathbb{M}} f(\bar{\bm{\phi}}_m^v) + \mu^v\mathsf{Pr}_{\bar{\bm{\phi}}_m}(\bm{\beta}^{v-1}),
\end{equation} 
where $\mu^v$ is the Riemannian version of the Polak-Ribi$\grave{\text{e}}$re formula \cite{boumal2014optimization}, which is given by
\begin{equation}
    \label{eq:mu}
    \mu^v = \frac{\left\langle\triangledown_{\mathbb{M}} f(\bar{\bm{\phi}}_m^{v}), \triangledown_{\mathbb{M}} f(\bar{\bm{\phi}}_m^{v})-\mathsf{Pr}_{\bar{\bm{\phi}}_m^v}(\triangledown_{\mathbb{M}} f(\bar{\bm{\phi}}_m^{v-1}))\right\rangle}{\langle\triangledown_{\mathbb{M}} f(\bar{\bm{\phi}}_m^{v-1}), \triangledown_{\mathbb{M}} f(\bar{\bm{\phi}}_m^{v-1})\rangle},
\end{equation}
where $\langle\mathbf{A},\mathbf{B}\rangle = \Re\{\mathsf{Tr}(\mathbf{A}^H\mathbf{B})\}$. 
The $v+1$-th update is given by a retraction function mapping point $\bar{\bm{\phi}}_m^v$ on the tangent space $\mathbb{T}_{\bar{\bm{\phi}}_m^v}\mathbb{M}$ to the manifold $\mathbb{M}$, that is \cite{boumal2014optimization}
\begin{equation}
    \label{eq:retraction}
    \bar{\bm{\phi}}_m^{v+1} = \frac{\bar{\bm{\phi}}_m^v + \varsigma^v\bm{\beta}^v}{\|\bar{\bm{\phi}}_m^v + \varsigma^v\bm{\beta}^v\|_2},
\end{equation}
where $\varsigma^v$ is the stepsize obtained by Armijo backtracking \cite{boumal2014optimization}.
The RCG procedure is summarized in Algorithm \ref{alg:RCG}.
 
% The computational complexity of the RCG algorithm mainly comes from steps 6 and 7, yielding a complexity bounded by $\mathcal{O}(L^{1.5})$.

\begin{algorithm}[!t]
    \caption{RCG: Riemannian Conjugate Gradient}
    \label{alg:RCG}
    \begin{algorithmic}[1]
        \REQUIRE $\nu_{\mathrm{c},k,m}$, $\nu_{\mathrm{p},l,m}$, $\chi_{\mathrm{c},k,m}$, $\chi_{\mathrm{p},l,m}$, $\xi_{\mathrm{c},k,m}$, $\xi_{\mathrm{p},l,m}$, $\forall k$, $\forall l$, $\bar{\bm{\phi}}_m$, $\varepsilon$, $m$, $\zeta$, $v_\mathrm{max}$.
        \ENSURE $\bar{\bm{\phi}}_m^\star$.
        \STATE{Initialize $\bar{\bm{\phi}}_m^0 = \bar{\bm{\phi}}_m$, $\bm{\beta}^0 = - \triangledown_{\mathbb{M}}f(\bar{\bm{\phi}}_m^0)$, $v = 0$.}
        \REPEAT 
            \STATE {Update $v = v + 1$.}
            \STATE {Calculate stepsize $\varsigma^{v-1}$ by Armijo backtracking \cite{boumal2014optimization}.}
            \STATE {Set $\bar{\bm{\phi}}_m^v$ by retraction (\ref{eq:retraction}).}
            \STATE {Update Polak-Ribi$\grave{\text{e}}$re formula $\mu^v$ by (\ref{eq:mu}).}
            \STATE {Update search direction $\bm{\beta}^v$ by (\ref{eq:xi}).}
        \UNTIL {$\|\triangledown_{\mathbb{M}}f(\bar{\bm{\phi}}_m^v)\|_2 \le \zeta$ or $v \ge v_\mathrm{max}$}
    \end{algorithmic}
\end{algorithm}

\textit{Summary:} With solutions to each cell of the multi-sector BD-RIS, the procedure of the  successive design is straightforward and summarized in Algorithm \ref{alg:design_phy}. 
The complexity of Algorithm \ref{alg:design_phy} is given by $\mathcal{O}(I_1L^{1.5})$, where $I_1$ is the number of iterations of Algorithm \ref{alg:design_phy}.

\begin{algorithm}[!t]
    \caption{Multi-Sector BD-RIS: Manifold Based Solution}
    \label{alg:design_phy}
    \begin{algorithmic}[1]
        \REQUIRE $\mathbf{X}_{\mathrm{c},k}$, $\bar{\mathbf{X}}_{\mathrm{p},l}$, $\mathbf{x}_{\mathrm{c},k}$, $\bar{\mathbf{x}}_{\mathrm{p},l}$, $\xi_{\mathrm{c},k}$, $\forall k$, $\bm{\phi}_l$, $\forall l$, $\varepsilon$, $\zeta$, $v_\mathrm{max}$.
        \ENSURE $\bm{\phi}_l^\star$, $\forall l$.
        \STATE{Initialize $\bm{\phi}_l^0 = \bm{\phi}_l$, $\forall l\in\mathcal{L}$, $v = 0$.}
        \REPEAT 
            \STATE {Update $v = v + 1$.}
            \FOR {$m\in\mathcal{M}$}
            \STATE {Update $\chi_{\mathrm{c},k,m}^v$, $\chi_{\mathrm{p},l,m}^v$, and $\xi_{k,m}^v$, $\forall k\in\mathcal{K}$ by (\ref{eq:aux}).}
            \STATE {Update $\bar{\bm{\phi}}_m^v$ by the RCG algorithm.}
            \ENDFOR
        \UNTIL {$\sum_{l}\|\bm{\phi}_l^v - \bm{\phi}_l^{v-1}\|_2^2 \le \zeta$ or $v\ge v_\mathrm{max}$}
    \end{algorithmic}
\end{algorithm}

\subsection{Algorithm and Analysis} 
\subsubsection{Framework of the Algorithm} 
The above three blocks are updated in an iterative manner until convergence as summarized in Algorithm \ref{alg:BCD}, where $\widehat{R}^v = \min_{\forall k}\widehat{R}_{\mathrm{c},k}^{v}+\sum_k\widehat{R}_{\mathrm{p},k}^{v}$.

\begin{algorithm}[!t]
    \caption{Joint Design: Block Coordinate Descent}
    \label{alg:BCD}
    \begin{algorithmic}[1]
        \REQUIRE $\mathbb{Q}_k$, $\sigma_k$, $R_{\mathrm{th},k}$, $\forall k$, $\forall l$, $P$, $\zeta$, $v_\mathrm{max}$.
        \ENSURE $\bm{\phi}_l^\star$, $\forall l$, $\mathbf{P}^\star$.
        \STATE{Initialize $\bm{\phi}_l^0$, $\bm{\mu}_l = \mathbf{0}$, $\forall l$, $\mathbf{P}^0$, $v = 0$.}
        \REPEAT 
            \STATE {Update $v = v + 1$.}          
            \STATE {Set weights $\{\bm{\lambda}_o^a, \forall o, \forall a\}^{v-1}$ by (\ref{eq:opt_weight}).}
            \STATE {Set equalizers $\{\mathbf{g}_o^{a}, \forall o, \forall a\}^{v-1}$ by (\ref{eq:opt_equalizer}).}
            \STATE {Update BD-RIS matrix $\{\bm{\phi}_l^v, \forall l\}$ by Algorithm \ref{alg:design_phy}.}
            \STATE {Update precoder $\mathbf{P}^v$ by solving SOCP (\ref{eq:sub_p1}).}
        \UNTIL {$\frac{|\widehat{R}^v - \widehat{R}^{v-1}|}{\widehat{R}^{v-1}} \le \zeta$ or $v\ge v_\mathrm{max}$}
    \end{algorithmic}
\end{algorithm}

\subsubsection{Initialization} 
In Algorithm \ref{alg:BCD}, $\bm{\phi}_l$, $\forall l\in\mathcal{L}$ are initialized as $\bm{\phi}_l^0 = \frac{1}{\sqrt{L}}[e^{\jmath\varphi_{l,1}},\ldots,e^{\jmath\varphi_{l,M}}]^T$, where $\varphi_{l,m}$ is randomly selected within the range $[0,2\pi]$,  $\forall l\in\mathcal{L}$, $\forall m\in\mathcal{M}$. The transmit power allocated to the common streams is initialized as $P_\mathrm{c} = 0.7P$, and thus that allocated to the private streams is $P_\mathrm{p} = 0.3P$. Accordingly, the common precoder is initialized as $\mathbf{p}_{\mathrm{c}} = \sum_{a}\sum_{k}\mathbf{Q}_k^a\bm{\phi}_l^*$, followed by a normalization $\mathbf{p}_{\mathrm{c}}^0 = \frac{\sqrt{P_\mathrm{c}}\mathbf{p}_{\mathrm{c}}}{\|\mathbf{p}_{\mathrm{c}}\|_2}$. The private precoder is initialized as $\mathbf{p}_{\mathrm{p},k} = \sum_{a}\mathbf{Q}_k^a\bm{\phi}_l^*$, followed by $\mathbf{p}_{\mathrm{p},k}^0 = \sqrt{\frac{P_\mathrm{p}}{K}}\frac{\mathbf{p}_{\mathrm{p},k}}{\|\mathbf{p}_{\mathrm{p},k}\|_2}$, $\forall k\in\mathcal{K}$.

\subsubsection{Complexity Analysis}
The complexity of Algorithm \ref{alg:BCD} mainly comes from the update of three blocks (steps 4-7). Specifically, updating block $\{\bm{\lambda}_o^a,\mathbf{g}_o^a, \forall o, \forall a\}$ requires complexity $\mathcal{O}(MNKLA)$. Updating block $\{\mathbf{P},\bar{\mathbf{c}}\}$ by solving SOCP has complexity $\mathcal{O}((NK+N+K)^{3.5})$, while updating block $\{\bm{\phi}_l,\forall l\}$ by Algorithm \ref{alg:design_phy} requires $\mathcal{O}(I_1L^{1.5})$. The overall complexity of Algorithm \ref{alg:BCD} is thus $\mathcal{O}(I(NK+N+K)^{3.5})$, where $I$ is the number of iterations of Algorithm \ref{alg:BCD} to achieve the convergence threshold.

\section{Performance Evaluation}
\label{sec:simulation}

In this section, we evaluate the performance of the proposed robust beamforming design for multi-sector BD-RIS aided RSMA under different CSI conditions. Particularly, we first illustrate the simulation settings. Then, we show convergence of the proposed algorithm. Finally, we evaluate the achievable ergodic sum-rate performance of multi-sector BD-RIS aided RSMA by analyzing the impact of the radiation pattern of BD-RIS antennas, CSI uncertainty, QoS threshold, and numbers of active/passive antennas. 

\subsection{Simulation Setup}

\begin{figure}[t]
    \centering
    \includegraphics[width = 0.48\textwidth]{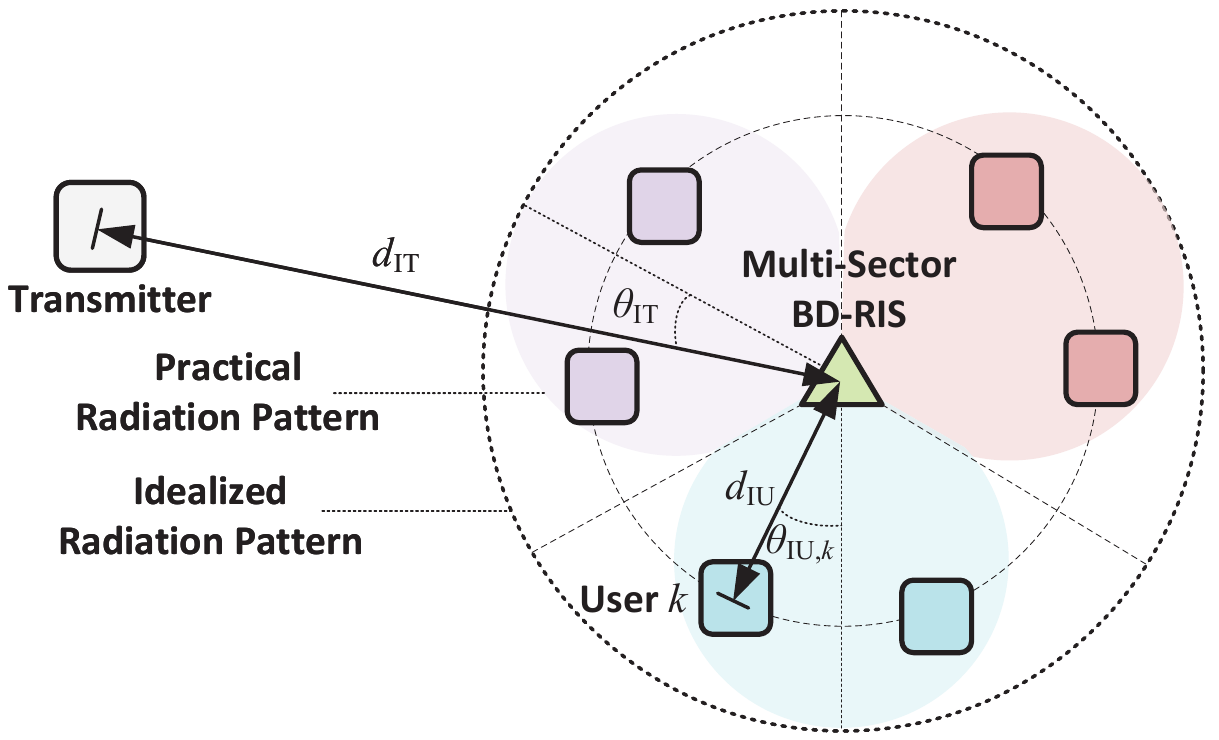}
    \caption{Top view of the locations among the transmitter, 3-sector BD-RIS, and users, and the illustration of idealized/practical radiation pattern.}\vspace{-0.2 cm}
    \label{fig:location}
\end{figure}

The perfect channels $\mathbf{Q}_k = \sqrt{\zeta_k^{-1}}\breve{\mathbf{G}}^H\mathsf{diag}(\breve{\mathbf{h}}_k)$, $\forall k\in\mathcal{K}$ are generated as a combination of small-scale fading and large-scale fading.
Specifically, the small-scale fading of transmitter-RIS and RIS-user channels is modeled as Rician fading, i.e.,
\begin{subequations}
    \begin{align}
    \breve{\mathbf{G}} &= \sqrt{\frac{\kappa}{\kappa + 1}}\breve{\mathbf{G}}^\mathrm{LoS} + \sqrt{\frac{1}{\kappa + 1}}\breve{\mathbf{G}}^\mathrm{NLoS}, \\
    \breve{\mathbf{h}}_k &= \sqrt{\frac{\kappa_k}{\kappa_k + 1}}\breve{\mathbf{h}}_k^\mathrm{LoS} + \sqrt{\frac{1}{\kappa_k + 1}}\breve{\mathbf{h}}_k^\mathrm{NLoS}, \forall k, 
    \end{align}
\end{subequations}
where $\{\kappa,\kappa_k,\forall k\}$ denotes the Rician factor, and channels with superscript ``LoS'' and ``NLoS'', respectively, denote line-of-sight (LoS) and non-LoS (NLoS) channels.
$\zeta_k$ is the path loss component measuring the large-scale fading. The large-scale fading components have different expressions when BD-RIS antennas have different radiation patterns. Here we consider two cases, that is each BD-RIS antenna has either idealized radiaton pattern or practical radiation pattern \cite{li2022beyond}, which are illustrated in Fig. \ref{fig:location} from the top view, yielding the following path loss model:
\begin{equation}
    \zeta_k = \begin{cases}
        \varrho (1-\cos\frac{\pi}{L})^2, & \text{Idealized Pattern},\\
        \frac{\varrho}{(\alpha_L+1)^2(\cos\theta_\mathrm{IT}\cos\theta_{\mathrm{IU},k})^{\alpha_L}}, & \text{Practical Pattern},
    \end{cases}\label{eq:path_loss}
\end{equation}
where $\varrho = \frac{4^3\pi^4d_\mathrm{IT}^{\epsilon_\mathrm{IT}}d_{\mathrm{IU},k}^{\epsilon_\mathrm{IU}}}{\lambda^4G_\mathrm{T}G_\mathrm{U}}$ is calculated by transmiter-RIS distance $d_\mathrm{IT}$, RIS-user distance $d_{\mathrm{IU},k}$, path loss exponents\footnote{The value of the path loss exponent is usually chosen within the range $[2,4]$ and depends on the small-scale fading conditions.} $\epsilon_\mathrm{IT}$ and $\epsilon_\mathrm{IU}$, wavelength of transmit signal $\lambda$, transmit antenna gain $G_\mathrm{T}$ and receive antenna gain $G_\mathrm{R}$. $\alpha_L$ is a parameter related to the beamwidth of BD-RIS antennas, $\theta_\mathrm{IT}$ is the elevation angle between the transmit antenna and BD-RIS antenna, and $\theta_{\mathrm{IU},k}$ is the elevation angle from the BD-RIS antenna to user $k$.
Sectors of the multi-sector BD-RIS are modeled as $M_\mathrm{x}\times M_\mathrm{y} = M$ uniform planar arrays and the transmitter is modeled as a uniform linear array with half-wavelength inter-antenna spacing. 
Locations among the transmitter, multi-sector BD-RIS, and users are shown in Fig. \ref{fig:location}.

The channel estimation error for each channel realization $\widetilde{\mathbf{Q}}_k$ is generalized such that $\mathsf{vec}(\tilde{\mathbf{Q}}_k)\sim\mathcal{CN}(\mathbf{0},\delta_k^2\mathbf{I}_{MN})$, where the channel estimation uncertainty $\delta^2_k$ is set as $\delta^2_k = \zeta_k^{-1}\delta^2$, $\forall k\in\mathcal{K}$ with a scalar $\delta\in[0,1)$. Specifically, the value $\delta = 0$ refers to the perfect CSI case.  
For each channel estimate $\widehat{\mathbf{Q}}_k = \mathbf{Q}_k - \widetilde{\mathbf{Q}}_k$, the $a$-th sample in the set (available at the transmitter) $\mathbb{Q}_k$ is given by $\mathbf{Q}^a_k = \widehat{\mathbf{Q}}_k + \widetilde{\mathbf{Q}}_k^a$, where $\widetilde{\mathbf{Q}}_k^a$ follows the same error distribution, that is $\mathsf{vec}(\widetilde{\mathbf{Q}}_k^a)\sim\mathcal{CN}(\mathbf{0},\delta_k^2\mathbf{I}_{MN})$,  $\forall k\in\mathcal{K}$, $\forall a\in\mathcal{A}$.
For each channel realization, the precoder and BD-RIS are designed based on sets $\mathbb{Q}_k$, $\forall k\in\mathcal{K}$ and the maximization of the average sum-rate as summarized in Algorithm \ref{alg:BCD}, while the sum-rate is calculated based on the perfect instantaneous channel $\mathbf{Q}$. 
The ergodic sum-rate is plotted as an average of sum-rates for 50 perfect channel realizations. 
The simulation settings, unless otherwise stated, are summarized in Table \ref{tab:simulate_set}.

\begin{table}[t] \caption{Simulation Settings}
    \centering
    \begin{tabular}{|c|c|}
        \hline
        Parameters & Value\\
        \hline
        \hline
        Rician Factor & $\kappa = \kappa_1 =\ldots=\kappa_K = 0$ dB\\
        \hline
        Transmit/Receive Antenna Gain & $G_\mathrm{t} = G_\mathrm{r} = 1$\\
        \hline
        Transmit Signal Frequency & $f = 2.4$ GHz\\
        \hline
        Transmitter-RIS Distance & $d_{\mathrm{TI}} = 100$ m\\
        \hline
        RIS-User Distance & $d_{\mathrm{IU},1} =\ldots= d_{\mathrm{IU},K} = 10$ m\\
        \hline
        Path Loss Exponent & $\epsilon_\mathrm{IT} = \epsilon_\mathrm{IU} = 2$\\
        \hline
        Transmitter-RIS Angle & $\theta_{\mathrm{IT}}\in[0,\frac{\pi}{L}]$\\
        \hline
        RIS-User Angle & $\theta_{\mathrm{IU},k}\in[0,\frac{\pi}{L}]$, $\forall k\in\mathcal{K}$\\  
        \hline
        Noise Power & $\sigma_1^2 =\ldots=\sigma_K^2 = -90$ dBm\\
        \hline
        Estimation Uncertainty & $\delta \in \{0,0.15\}$\\
        \hline
        No. of Samples & $A = 50$\\
        \hline
        No. of Sectors & $L = 3$\\
        \hline
        No. of Users per Sector & $K_1 = \ldots=K_L = 2$\\
        \hline
        No. of BD-RIS antennas per Sector & $M = M_\mathrm{x}\times M_\mathrm{y} = 5\times 4 = 20$\\
        \hline
        No. of transmit antennas & $N = 4$\\
        \hline
        QoS Threshold per User & $R_{\mathrm{th},1} = \ldots = R_{\mathrm{th},K} = R_\mathrm{th}$\\
        \hline
    \end{tabular}\label{tab:simulate_set}
\end{table}

\begin{figure}
    \centering
    \includegraphics[width=0.48\textwidth]{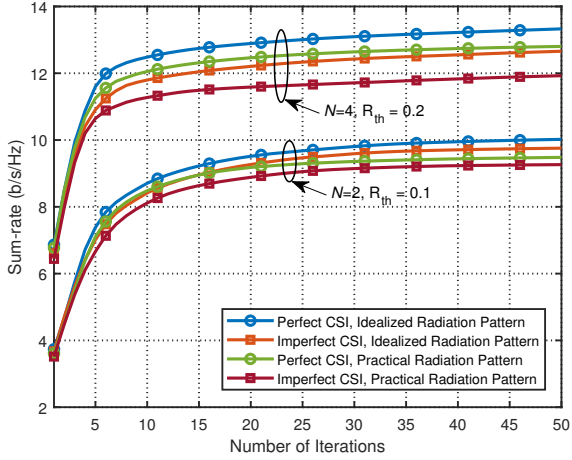}
    \caption{Sum-rate versus the number of iterations ($P = 35$ dBm).}
    \vspace{-0.2 cm}\label{fig:ASR_Iter}
\end{figure}

\subsection{Convergence of the Algorithm}
The convergence of Algorithm \ref{alg:BCD} cannot be theoretically proved since the design of BD-RIS matrix block does not lead to a global optimal solution. Instead, we provide numerical results to show the proposed algorithm can converge within limited iterations.
Fig. \ref{fig:ASR_Iter} plots the sum-rate versus the number of iterations of Algorithm \ref{alg:BCD} for different CSI conditions and parameter settings.
We can observe from Fig. \ref{fig:ASR_Iter} that the proposed algorithm always converges within finite iterations regardless of simulation settings, such as channel conditions (perfect and imperfect CSI), the number of transmit antennas, and QoS requirements.

\begin{figure}
    \centering
    \includegraphics[width=0.48\textwidth]{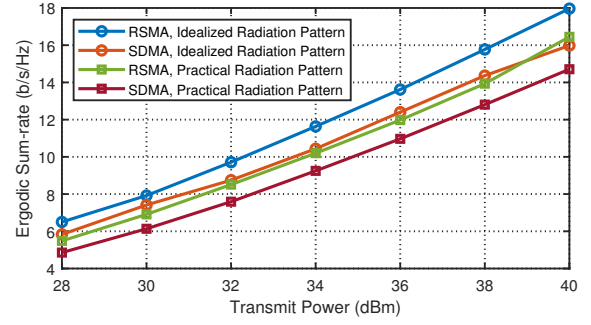}
    % \subfigure[Imperfect CSI]{
    % \includegraphics[width=0.49\textwidth]{figures/ESR_P_ImPerfectCSI.eps}}
    \caption{Ergodic sum-rate versus transmit power with perfect CSI ($R_\mathrm{th} = 0.2$).}\vspace{-0.2 cm}
    \label{fig:ESR_P}
\end{figure}

\subsection{Impact of the Radiation Pattern of BD-RIS Antennas}

We first evaluate the ergodic sum-rate of multi-sector BD-RIS aided RSMA when BD-RIS antennas have either idealized or practical radiation patterns in Fig. \ref{fig:ESR_P}. For comparison, we add the performance of multi-sector BD-RIS assisted SDMA\footnote{In this work, we do not involve the NOMA scheme in the simulations since the benefits of RSMA compared to NOMA have been widely investigated in \cite{mao2019rate,mao2022rate,mishra2021rate} and that in multi-antenna systems RIS aided SDMA is a much tighter benchmark than the NOMA scheme \cite{li2022rate}.}, which is achieved by switching off the common precoder at the transmitter\footnote{In this case, the multi-sector BD-RIS design with given precoder is performed by using the closed-form solution proposed in \cite{li2022beyond}.}. We have the following observations. 1) With the same multiple access technique, the idealized radiation pattern scheme always outperforms the practical radiation pattern scheme, which is in accordance with the results in \cite{li2022beyond}. 2) With the same radiation pattern for each BD-RIS antenna, the RSMA scheme always outperforms the SDMA scheme, which demonstrates the benefits of integrating RSMA and multi-sector BD-RIS. 
3) The BD-RIS aided RSMA with practical radiation pattern of BD-RIS antennas achieves similar performance to the BD-RIS aided SDMA with idealized radiation pattern for BD-RIS antennas. 
This phenomenon demonstrates the benefits of integrating RSMA and multi-sector BD-RIS in easing the requirements to antenna realizations. 

\begin{figure}
    \centering
    \includegraphics[width=0.48\textwidth]{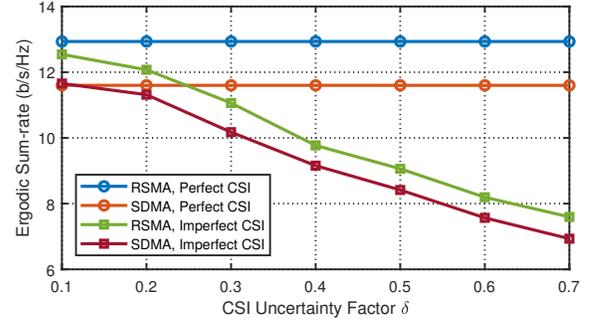}
    \caption{Ergodic sum-rate versus CSI uncertainty $\delta$ with idealized radiation pattern ($P = 35$ dBm, $R_\mathrm{th} = 0.2$).}\vspace{-0.2 cm}
    \label{fig:ESR_delta}
\end{figure}

\subsection{Impact of CSI Uncertainty}

In Fig. \ref{fig:ESR_delta}, we evaluate the impact of CSI uncertainty by plotting the ergodic sum-rate versus $\delta$ when each BD-RIS has an idealized radiation pattern. We observe from Fig. \ref{fig:ESR_delta} that the ergodic sum-rate for both RSMA and SDMA schemes decreases with increasing channel estimation errors. 
However, the multi-sector BD-RIS aided RSMA always outperforms multi-sector BD-RIS aided SDMA regardless of CSI conditions, which demonstrates that integrating RSMA and multi-sector BD-RIS always provides performance gain over SDMA schemes even though with limited CSI.

\begin{figure}
    \centering
    \subfigure[Ergodic sum-rate versus $R_\mathrm{th}$]{
    \includegraphics[width=0.48\textwidth]{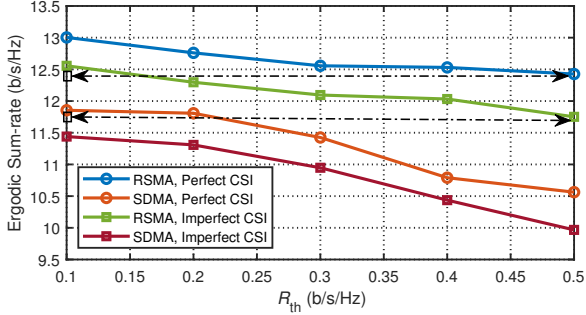}}
    \subfigure[Power allocation versus $R_\mathrm{th}$]{
    \includegraphics[width=0.48\textwidth]{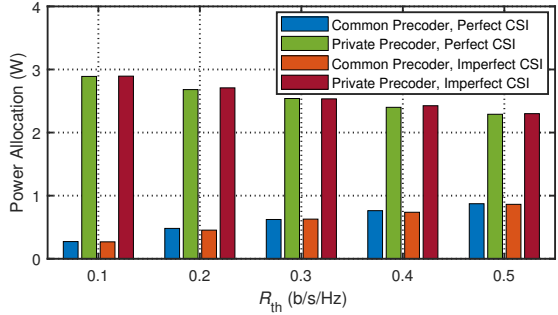}}
    \caption{Impact of QoS threshold with idealized radiation pattern and different CSI conditions ($P = 35$ dBm).}
    \label{fig:ESR_Rth}
\end{figure}

\subsection{Impact of QoS Threshold}

Then, in Fig. \ref{fig:ESR_Rth}, we show the impact of the QoS requirement for each user to the achievable ergodic sum-rate under different CSI conditions when each multi-sector BD-RIS antenna has an idealized radiation pattern. From Fig. \ref{fig:ESR_Rth} we have the following observations. 1) The multi-sector BD-RIS aided RSMA always outperforms the SDMA scheme with various QoS thresholds and different CSI conditions. 2) The achievable ergodic sum-rate for both schemes decreases with the growth of the QoS requirement, which can be explained as follows. Since we focus on the sum-rate maximization problem, when the number of transmit antennas is smaller than the number of users ($4 = N < K = 6$), the weakest $N-K$ users tend to be turned off. In this case, increasing the QoS requirement for all users violates the aim of maximizing sum-rate. 
3) The performance gap between the RSMA scheme and the SDMA scheme becomes larger with increasing QoS threshold for each user, and thus the RSMA scheme is able to achieve both higher sum-rate and better QoS than the SDMA case, e.g., the comparison between RSMA scheme with $R_\mathrm{th} = 0.5$ and SDMA scheme with $R_\mathrm{th} = 0.1$.  
This is because in RSMA scheme, the power allocated to the common precoders increases with the QoS requirement. 
This phenomenon demonstrates that the common precoders play an increasingly important role to guarantee the fairness while achieving better sum-rate performance.

\begin{figure}
    \centering
    \includegraphics[width=0.48\textwidth]{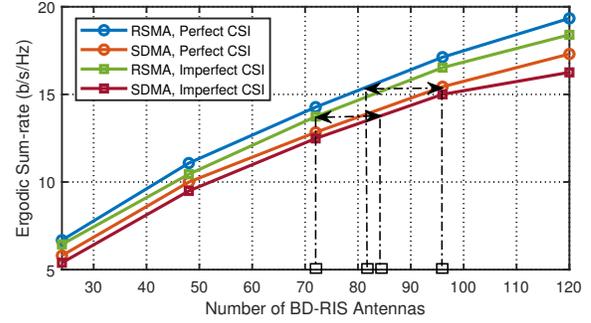}
    \caption{Ergodic sum-rate versus the total number of antennas of BD-RIS with idealized radiation pattern ($P = 35$ dBm, $R_\mathrm{th} = 0.2$).}
    \label{fig:ESR_M}
\end{figure}

\begin{figure}
    \centering
    \includegraphics[width=0.48\textwidth]{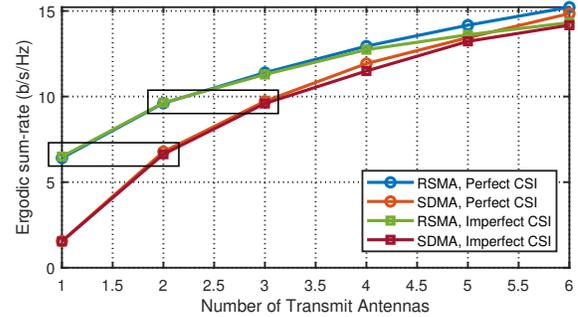}
    \caption{Ergodic sum-rate versus the number of transmit antennas when BD-RIS antennas have idealized radiation pattern ($P = 35$ dBm, $R_\mathrm{th} = 0.15$).}
    \label{fig:ESR_N}
\end{figure}

\subsection{Impact of Numbers of Active and Passive Antennas}

In Fig. \ref{fig:ESR_M}, we plot the ergodic sum-rate versus the total number of antennas of the multi-sector BD-RIS under different CSI conditions when each BD-RIS antenna has an idealized radiation pattern. We observe from Fig. \ref{fig:ESR_M} that the ergodic sum-rate of both RSMA and SDMA schemes grows with increasing number of BD-RIS antennas, and that the RSMA scheme always achieves better performance than the SDMA scheme. More importantly, to achieve the same sum-rate, the required number of passive antennas can be reduced compared to the SDMA scheme by adopting RSMA at the transmitter. 
For example, in perfect CSI case, to achieve an ergodic sum-rate of 15 b/s/Hz, the required number of passive antennas can be reduced by 16\%; in imperfect CSI case, to achieve an ergodic sum-rate of 13 b/s/Hz, the required number can be reduced by 14\%. This observation highlights the benefit of synergizing BD-RIS and RSMA in saving on passive antennas, which is friendly to the hardware implementation of multi-sector BD-RIS.

Finally, in Fig. \ref{fig:ESR_N}, we plot the ergodic sum-rate versus the number of transmit antennas under different CSI conditions. 
Similar conclusions that the RSMA scheme always outperforms the SDMA scheme can be obtained from Fig. \ref{fig:ESR_N}. 
More importantly, to achieve the same sum-rate, the required number of active antennas at the transmitter can also be reduced compared to the SDMA scheme when RSMA is adopted at the transmitter. For example, an 1-antenna RSMA-enabled transmittersupports the same sum-rate performance as an 2-antenna SDMA-enabled transmitter. 
Such reduction of the number of transmit antennas demonstrates the benefit of synergizing BD-RIS and RSMA in reducing power consumption and cost, which leads to an energy efficiency gain.

\section{Conclusion}
\label{sc:Conclusion}

In this work, we consider the integration of multi-sector BD-RIS and RSMA in a MU-MISO communication system. 
Specifically, we assume a multi-sector BD-RIS and an RSMA-enabled transmitter. 
Given that the channel estimation for RIS and BD-RIS aided scenarios is challenging, we model the channel estimation error when only partial/imperfect CSI is available at the transmitter. 

With the multi-sector BD-RIS aided RSMA model and CSI uncertainty model, we propose a robust design to jointly optimize the transmit precoder and multi-sector BD-RIS for a MU-MISO system. Specifically, SAA and WMMSE-rate relationship are adopted to transform the stochastic problem into a more tractable deterministic multi-block problem. Regarding the design of the BD-RIS matrix block, we adopt the RCG algorithm after transforming the corresponding problem into a smooth optimization.

Finally, we present simulation results, which demonstrate the performance enhancement of multi-sector BD-RIS aided RSMA compared to multi-sector BD-RIS aided SDMA with different radiation pattern of BD-RIS antennas and various parameter settings, such as CSI conditions and QoS threshold for users. 
More importantly, with given sum-rate requirement, the numbers of active and passive antennas can be effectively reduced by synergizing RSMA with multi-sector BD-RIS. Such reduction of the number of antennas is beneficial for practical implementations. 

Future research avenues include, but are not limited to the following aspects:

\textit{1) Investigating User Selection for Multi-Sector BD-RIS:} It is interesting to consider the BD-RIS aided communication system with mobile users. In this case, imperfect CSI may lead to inaccurate selection between users and sectors of BD-RIS, such that new metrics and algorithms should be developed to effectively involve the impact of user selection.

\textit{2) Developing Different Multi-Sector BD-RIS Design Algorithms:} While the performance gain of multi-sector BD-RIS aided RSMA over SDMA schemes has been evaluated by the proposed algorithm in this work, it would also be interesting to develop other algorithms with lower complexity, better performance, and better robustness to the CSI uncertainty.

\textit{3) Modeling the Inter-Sector Interference of Multi-Sector BD-RIS:} In this work, we assume each BD-RIS antenna has a perfect uni-directional radiation pattern such that there are no overlaps between different sectors of BD-RIS. However, in practical implementations, the BD-RIS antennas might have strong sidelobes causing overlaps, such that users will be covered by more than one sector of the BD-RIS. When taking into account this effect, the beamforming design and analysis based on the perfect model do not hold and thus new insights should be investigated for imperfect BD-RIS models.

\bibliographystyle{IEEEtran}
\bibliography{references}

\end{document}